%% file: ms.tex
\documentclass[a4paper,openany, 12pt]{book}

\usepackage[text={15.5cm,23cm},centering]{geometry}

\newcommand\savemathcal[1]{%
  \expandafter\newsavebox\csname mc#1content\endcsname%
  \expandafter\savebox\csname mc#1content\endcsname{$\mathcal{#1}$}%
  \expandafter\newcommand\csname mc#1\endcsname{%
    \expandafter\usebox\expandafter{\csname mc#1content\endcsname}}%
}
\newcommand\altmathcal[1]{\csname mc#1\endcsname}
\savemathcal{Z}
\savemathcal{B}

\usepackage{mathptmx}
\usepackage{graphicx}
\usepackage{chapterbib} 
\usepackage{subfigure}
\usepackage{color}
\usepackage{amsmath}
\usepackage{amssymb}
\usepackage[colorlinks=true, linkcolor=blue]{hyperref}
\usepackage[english]{babel}  

\title{The cosmic 21-cm revolution: charting the first billion years of our Universe}
\author{Andrei Mesinger}

\begin{document}



\include{Bernardi/chapter}



\end{document}

%% file: Bernardi/chapter.tex
\setcounter{chapter}{4}
\chapter{21~cm observations: calibration, strategies, observables}
\label{chapter:bernardi}

\begin{bf}
Gianni Bernardi (INAF - IRA \& Rhodes University) \\
  
Abstract\\

This chapter aims to provide a review of the basics of 21~cm interferometric observations and its methodologies. A summary of the main concepts of radio interferometry and their connection with the 21~cm observables - power spectra and images - is presented. I then provide a review of interferometric calibration and its interplay with foreground separation, including the current open challenges in calibration of 21~cm observations. Finally, a review of 21~cm instrument designs in the light of calibration choices and observing strategies follows.
\end{bf}

\section{Interferometry overview}

\begin{figure}[]
\begin{center}
\includegraphics[width=1.\textwidth]{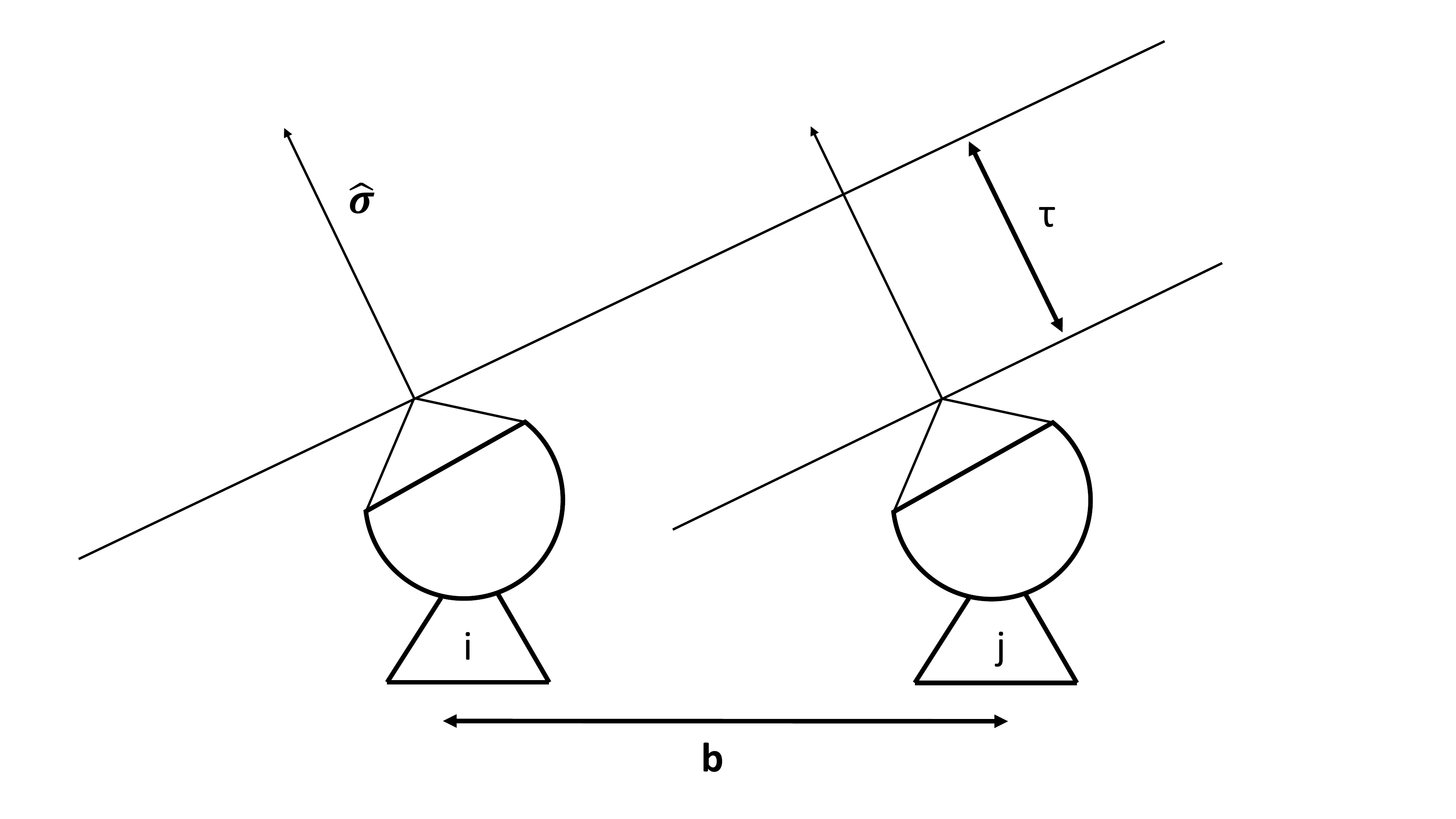}
\end{center}
\caption{A standard schematic of the two element interferometer.}
\label{fig:fig1}
\end{figure}

The Van Cittert-Zernike theorem expresses the fundamental relationship between the sky spatial brightness (or brightness distribution) $I$ and the quantity measured by an interferometer, i.e. the visibility $V$ (e.g., \cite{TMS}):
\begin{equation}
V_{ij} ({\bf b}, \nu) = \int_\Omega {\bar I} (\hat{\bf \sigma}, \nu) \, e^{-2 \pi i \nu \frac{{\bf b} \cdot {\hat {\bf \sigma}}}{c}} d \hat{\bf \sigma},
\label{eq:1}
\end{equation}
where ${\bf b}$ is the baseline vector that separates antenna~$i$ and antenna~$j$, $\nu$ is the observing frequency, $\hat{\sigma}$ is the observing direction (see Figure~\ref{fig:fig1}), $c$ the speed of light and the integral is taken over the source size $\Omega$. 
It can be seen in Figure~\ref{fig:fig1} that the celestial signal travels an extra path between the two antennas, and that length corresponds to a geometrical time delay $\tau = \frac{{\bf b} \cdot {\hat {\bf \sigma}}}{c}$, where the word ``geometrical" refers to the fact that the delay depends upon the source position in the sky and the relative separation between the two antennas. Equation~\ref{eq:1} can be derived as the output of the correlator, the digital equipment responsible to combine signals from antenna pairs. The voltage induced by a celestial source at any antenna can be written in a generic form as $(v \cos{2\pi \nu t})$, and the correlator will output the time average product of the voltages $r$ measured by antenna $i$ and $j$ respectively:
\begin{eqnarray}
r & = & \langle v_i(t)v_j(t) \rangle = \langle v^2 \cos{(2\pi \nu t)} \cos{[2\pi \nu (t - \tau)]} \rangle \nonumber \\
			     & = & \langle v^2 \frac{\cos{(2\pi \nu \tau)} + \cos{(4\pi \nu t - 2\pi \nu \tau)}}{2} \rangle. 	
\label{eq:1.1}
\end{eqnarray}
While the first term of the right hand side of equation~\ref{eq:1.1} varies slowly with the Earth rotation, the second oscillates rapidly for any typical radio observations ($\nu > 10$~MHz) and averages to zero, leading to:  
\begin{equation}
r(\tau) \approx \frac{v^2}{2} \cos{2\pi \nu \tau},
\label{eq:1.2}
\end{equation}
which is a sinusoidal pattern termed ``fringe".
Equation~\ref{eq:1.2} actually represents the contribution to the fringe from the pointing direction. We can obtained the contribution from the whole source by integrating over the source size and adding the odd (sine) to the even (cosine) fringe component to form a general, complex-valued fringe $R$: 
\begin{equation}
R ({\bf b}, \nu) = \int_\Omega r(\tau) d\tau = \int_\Omega \frac{v^2}{2} (\cos{2\pi \nu \tau} - i\sin{2\pi \nu \tau}) = \int_\Omega \frac{v^2}{2} e^{-2 \pi i \nu \frac{{\bf b} \cdot {\hat {\bf \sigma}}}{c}} d \hat{\bf \sigma},
\label{eq:1.3}
\end{equation}
where I have substituted the definition of geometrical delay in the last step. If we note that the $\frac{v^2}{2}$ voltage square term depends upon the direction in the sky as it is proportional to the source brightness, equation~\ref{eq:1.3} is essentially equivalent to equation~\ref{eq:1} and shows how the correlator outputs directly the spatial coherence function of the sky emission, i.e. the visibility.  

The sky brightness distribution $I$ does not appear directly in the Van Cittert-Zernike theorem, but filtered by the antenna primary beam response $A$ that depends upon the direction in the sky and the wavelength, i.e. ${\bar I} (\hat{\bf \sigma}, \nu) = A (\hat{\bf \sigma}, \nu)  \, I (\hat{\bf \sigma}, \nu)$. The response of the primary beam attenuates the sky emission away from the pointing direction, effectively reducing the field of view $\Omega_F$ of the instrument. Generally speaking, the size of the field of view is essentially given by the antenna diameter $D$: 
\begin{equation}
\Omega_F \approx \frac{\lambda}{D},
\label{eq:2}
\end{equation}
where $\lambda$ is the observing wavelength.

\begin{figure}[]
\begin{center}
\includegraphics[width=1.\textwidth]{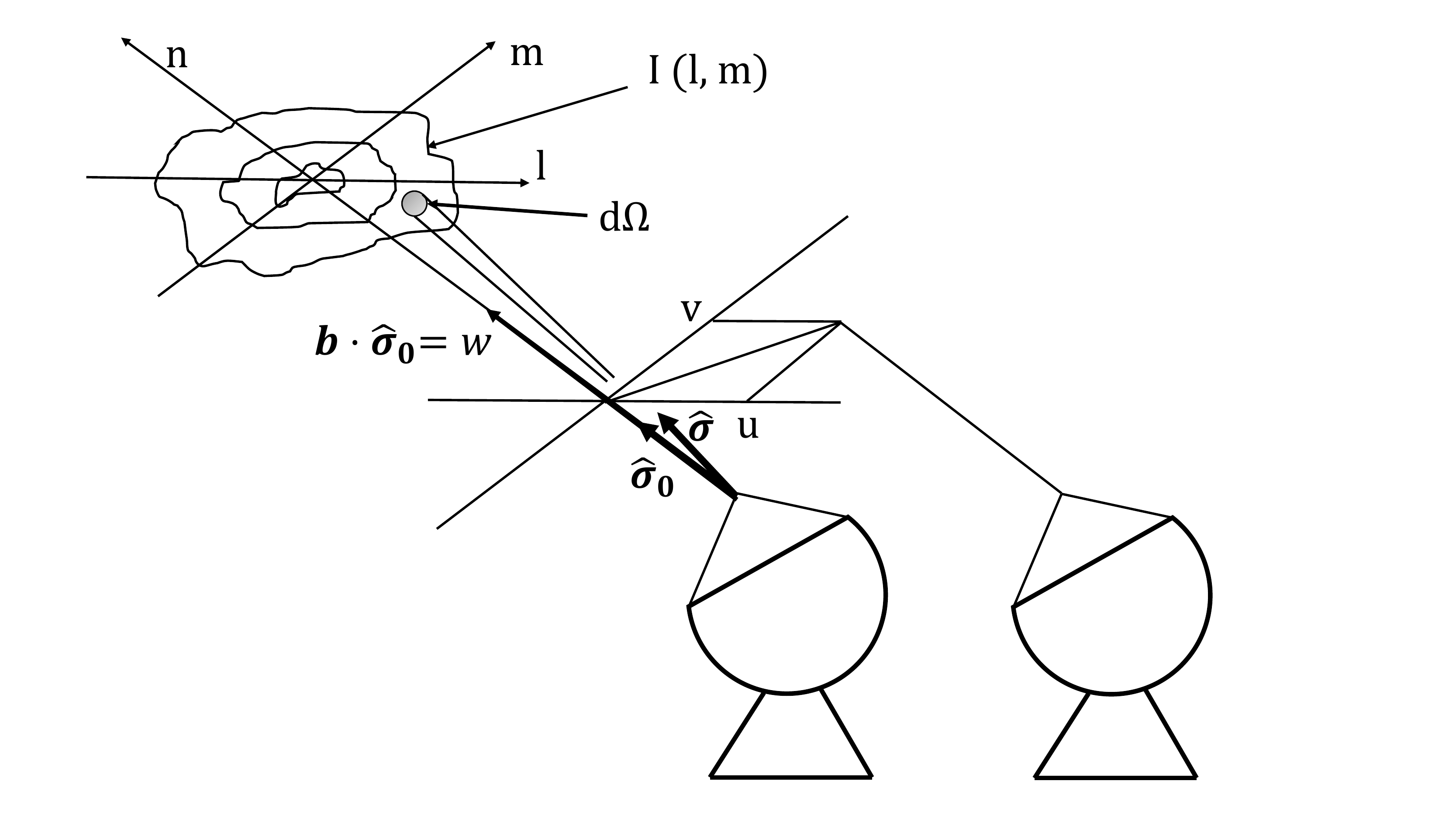}
\end{center}
\caption{Cartoon representation of the coordinate system used for interferometric imaging. The $(l,m)$ plane is tangent to the sky.}
\label{fig:fig1b}
\end{figure}

The Van Cittert-Zernike theorem that defines the visibility function is often re-written in a different coordinate system, i.e. using the components of the baseline vector $(u,v,w)$, where $(u,v)$ are the components of the baseline vector in the plane of the array and $w$ is the component along the pointing direction $\sigma_0$ (Figure~\ref{fig:fig1b}). The sky position in the $\hat {\sigma}$ direction can be decomposed into the $(l,m)$ components parallel to the plane of the sky and the $n$ component along the $w$ axis. In this system, coordinates can be re-written as (\cite{TMS}):
\begin{eqnarray}
\frac{\nu \, {\bf b} \cdot {\hat {\bf \sigma}}}{c} & = & ul + vm + wn, \nonumber \\
\frac{\nu \, {\bf b} \cdot {\hat {\bf \sigma}}_0}{c} & = & w, \nonumber \\
d\Omega & = & \frac{dl dm}{n} = \frac{dl dm}{\sqrt{1 - l^2 - m^2}}
\label{eq:2.1}
\end{eqnarray}
and equation~\ref{eq:1} then becomes:
\begin{equation}
V_{ij} (u,v,w, \nu) = \int_\Omega {\bar I} (l, m, \nu) \, e^{-2 \pi i (ul + vm + w(\sqrt{1 - l^2 - m^2} - 1))} \frac {dl \, dm \, dn}{\sqrt{1 - l^2 - m^2}}.
\label{eq:3}
\end{equation}
Although low frequency radio observations are intrinsically wide-field, for the purpose of studying the 21~cm observables, we can reduce equation~\ref{eq:3} to a two dimensional Fourier transform:
\begin{equation}
V_{ij} (u,v, \nu) = \int_\Omega {\bar I} (l, m, \nu) \, e^{-2 \pi i (ul + vm)} dl \, dm.
\label{eq:4}
\end{equation}
Equation~\ref{eq:4} indicates that an {\it interferometer measures the two dimensional Fourier transform of the spatial sky brightness distribution}. If our goal is to reconstruct the sky brightness distribution, equation~\ref{eq:4} can be inverted into its corresponding Fourier pair:
\begin{equation}
{\bar I} (l, m, \nu) = \int V_{ij} (u,v, \nu) \, e^{2 \pi i (ul + vm)} du \, dv.
\label{eq:5}
\end{equation}
Equation~\ref{eq:5} is, however, a poor reconstruction of the sky brightness distribution as only one Fourier mode is sampled at a single time instance. Strictly speaking, indeed, all the quantities in equation~\ref{eq:4} and \ref{eq:5} are time variable. In most cases, the time dependence of the primary beam and the sky brightness distribution can be neglected, however, this is not the case for the visibility $V$ as the projection of the baseline vector with respect to the source direction changes significantly throughout a long (e.g. a few hours) track. In this way, many measurements of the visibility coherence function $V$ can be made as $(u,v)$ change with time, allowing for a better reconstruction of the ${\bar I} (l, m, \nu)$ function. This method is commonly referred to as {\it filling the $uv$ plane via Earth rotation synthesis} and was invented by \cite{ryle60}. The other (complementary) way to fill the $uv$ plane is to deploy more antennas on the ground in order to increase the number of instantaneous measurements of independent Fourier modes. If $N$ antennas are connected in an interferometric array, $\frac{N (N - 1)}{2}$ instantaneous measurements are made. 

The combination of a large number of antennas and the Earth rotation synthesis, defines the sampling function $S(u,v)$ in the $uv$ plane. In any real case, equation~\ref{eq:5} can therefore be re-written as:
\begin{equation}
{\bar I}_D (l, m, \nu) = \int S(u,v, \nu) V (u,v, \nu) \, e^{2 \pi i (ul + vm)} du \, dv,
\label{eq:6}
\end{equation}
where ${\bar I}_D$ indicates the sky brightness distribution sampled at a finite number of $(u,v)$ points (often termed {\it dirty image}) and where the explicit dependence on the antenna pair was dropped for simplicity. Using the convolution theorem, equation~\ref{eq:6} can be re-written as:
\begin{equation}
{\bar I}_D (l, m, \nu)  =  \tilde{S \, V} =  {\tilde S} \ast {\tilde V} = {\rm PSF} (l, m, \nu) \ast {\bar I} (l, m, \nu),
\label{eq:7}
\end{equation}
where the tilde indicates the Fourier transform, $\ast$ the convolution operation and PSF is the Point Spread Function, i.e. the response of the interferometric array to a point sources which, in our case, is also the Fourier transform of the $uv$ coverage.

The sampling function always effectively reduces the integral over a finite (often not contiguous) area of the $uv$ plane. In particular, the sampled $uv$ plane is restricted between a minimum $uv$ distance that cannot be shorter than the antenna\footnote{In this chapter I use the words ``antenna" and ``station" interchangeably to indicate the correlated elements even if, in the literature, they are normally used to indicate a dish and a cluster of dipoles, respectively.} size and the largest separation between antennas, i.e. the maximum baseline ${\bf b}_{\rm max}$. The maximum baseline also sets the maximum angular resolution $\theta_b$:
\begin{equation}
\theta_b \approx \frac{\lambda}{|{\bf b}_{\rm max}|}.
\label{eq:8}
\end{equation}
The incomplete sampling of the $uv$ space leads to a PSF that has ``sidelobes", i.e. nulls and secondary lobes that can often contaminate fainter true sky emission. The best reconstruction of the sky brightness distribution ${\bar I}$ requires deconvolution of the dirty image from the PSF. 

\begin{figure}[]
\begin{center}
\includegraphics[width=1.\textwidth]{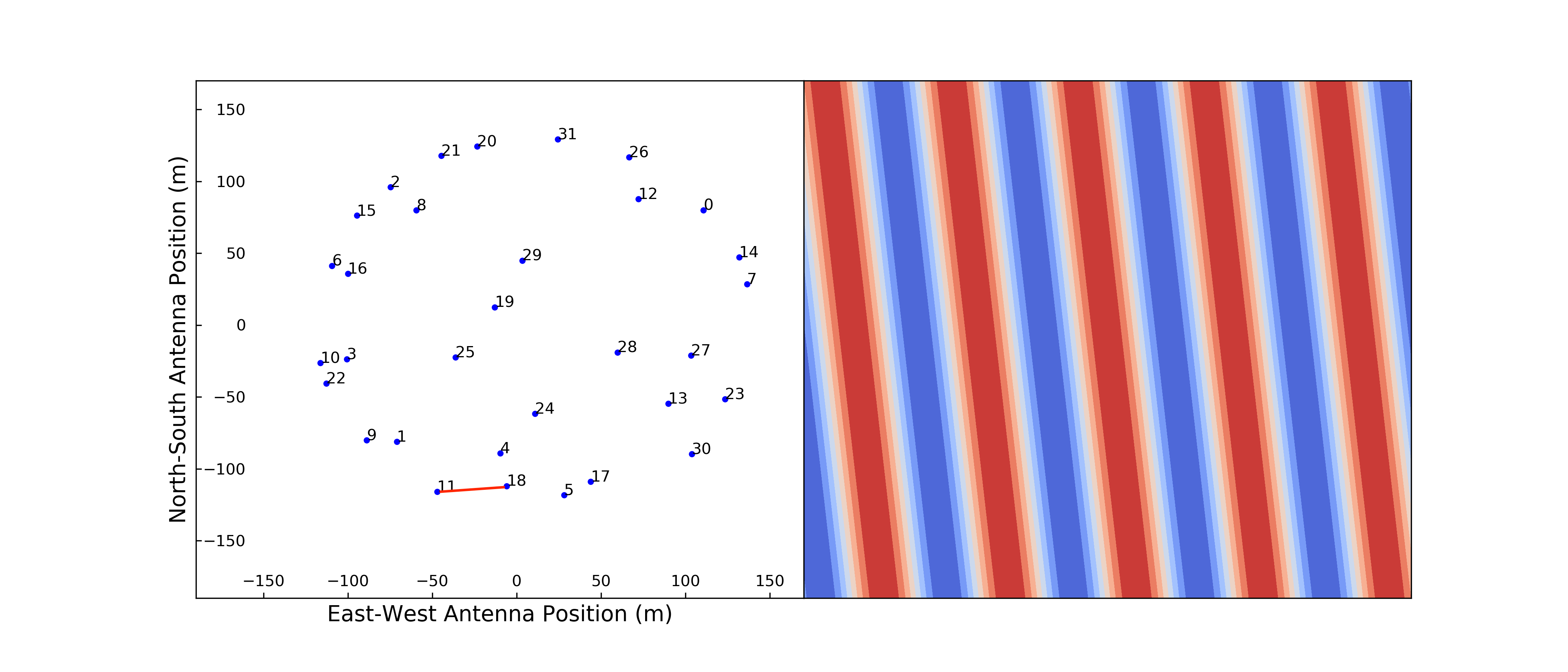}
\includegraphics[width=1.\textwidth]{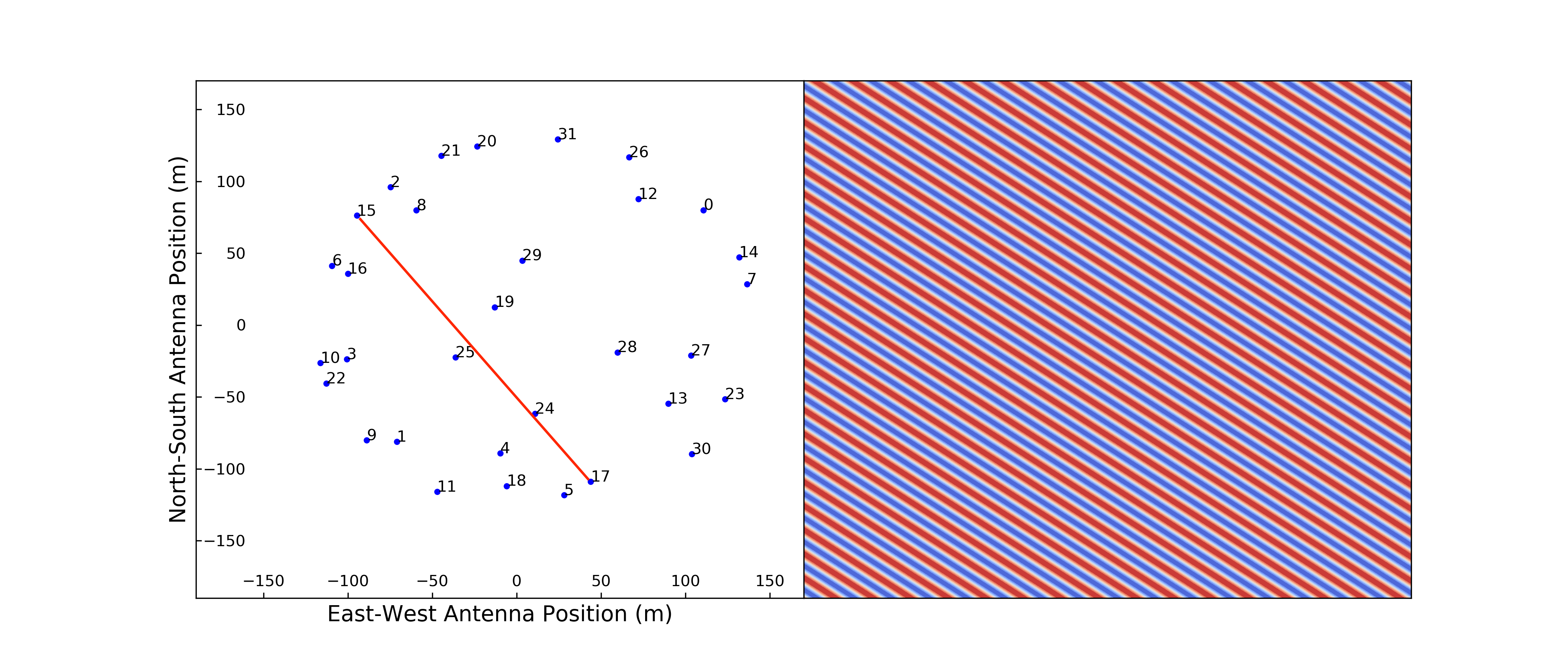}
\end{center}
\caption{Left panel: one baseline selected (indicated with a red line) from 32 antennas distributed within a 350~m circle (taken from \cite{jacobs11}). Right panel: corresponding point spread function. For a single baseline case, the point spread function is essentially a sinusoidal fringe pattern whose period is inversely proportional to the baseline length, i.e. the pattern correponding to $\sim 50$~m baseline (top) oscillates approximately seven times slower than a $\sim 350$~m baseline (bottom). The fringe phase (i.e. the pattern orientation) is given by the baseline orientation.}
\label{fig:fig1c}
\end{figure}

\begin{figure}[]
\begin{center}
\includegraphics[width=1.\textwidth]{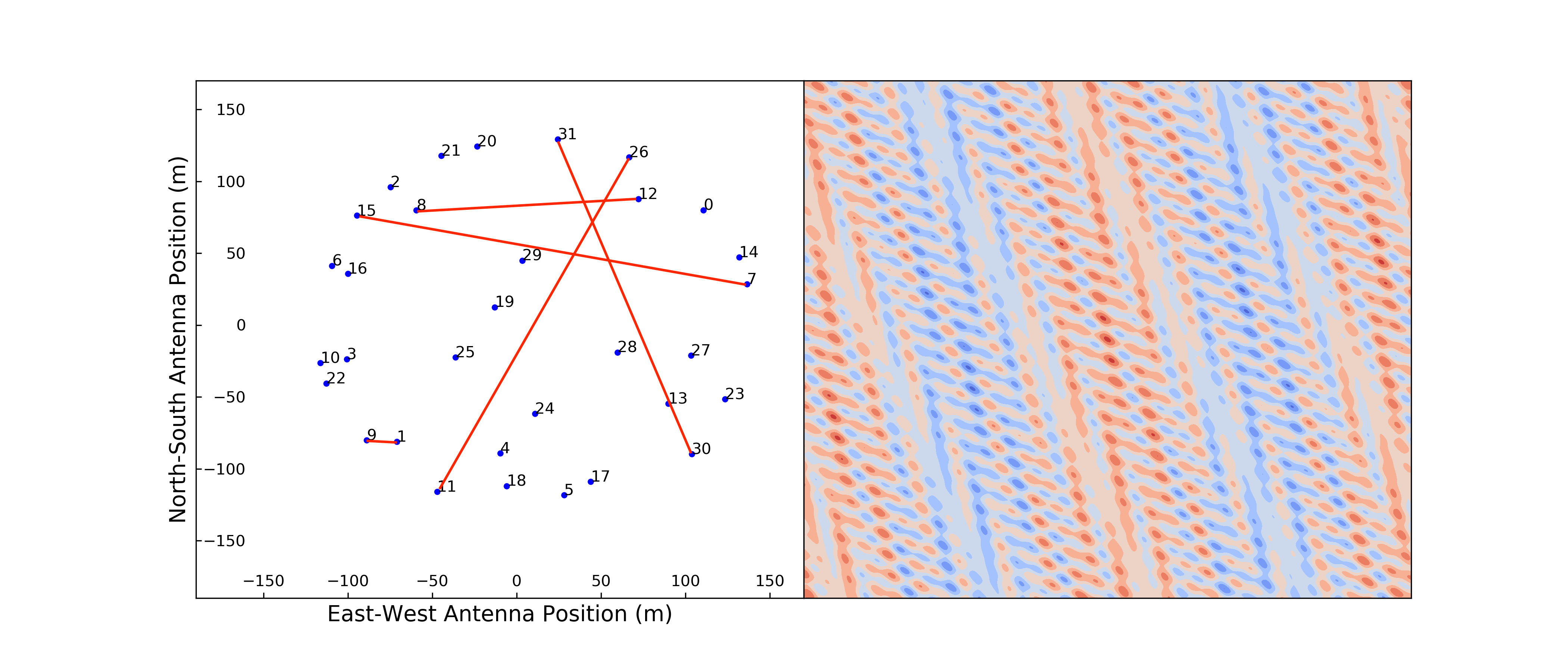}
\includegraphics[width=1.\textwidth]{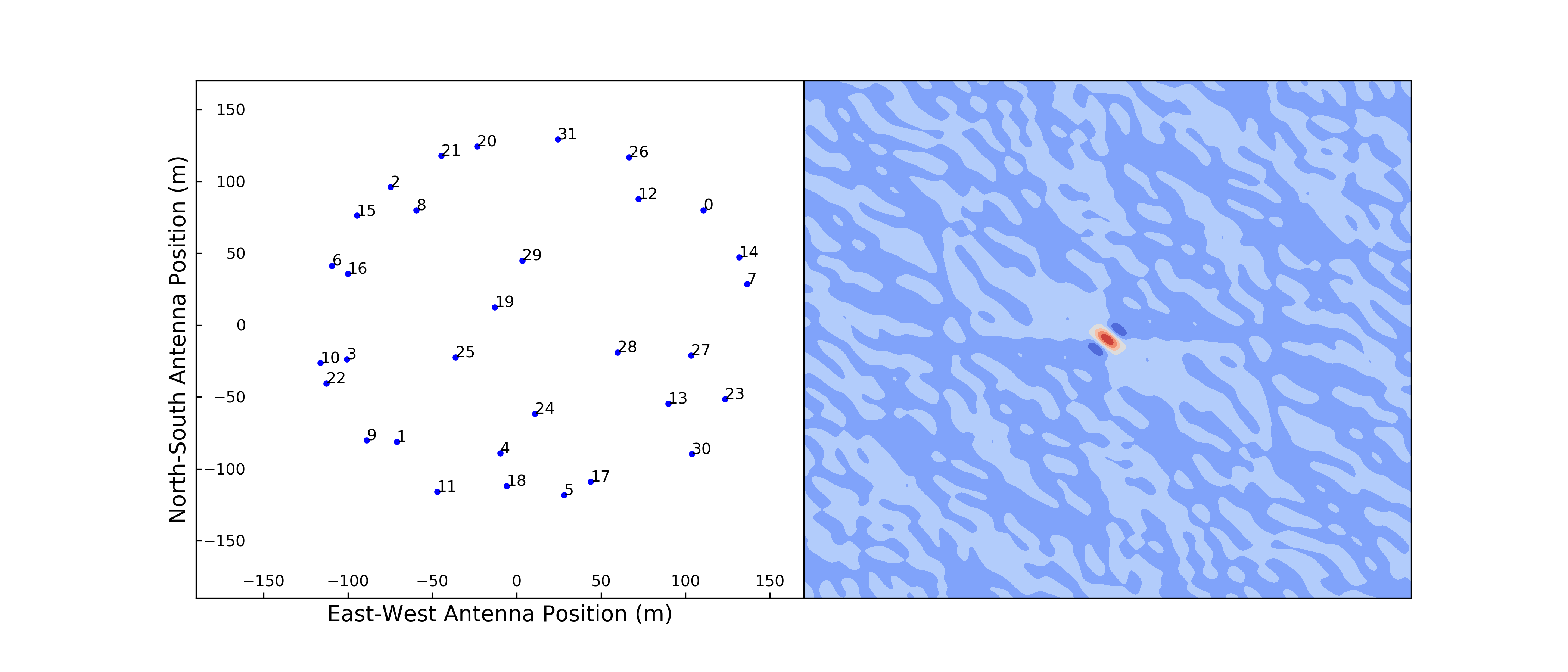}
\end{center}
\caption{Same as Figure~\ref{fig:fig1c}, but including five baselines with different lengths and orientations (red lines, top panel) and all the baselines (for $N = 32$ there are 496 baselines; bottom panel) simultaneously. The fringe pattern is already noticeably different when five baselines are included with respect to the single baseline, although a clean point spread function only appears when all the baselines are used simultaneously (bottom panel).}
\label{fig:fig1d}
\end{figure}

Figure~\ref{fig:fig1c} and \ref{fig:fig1d} provide an example of the sampling $S(u,v)$ and the corresponding point spread function. A single baseline essentially imprints a (sinusoidal) fringe pattern on the sky, whose period and phase depend upon the baseline length and orientation respectively (equation~\ref{eq:1}). The combination of more baselines of different lengths and orientations improves the sampling function until a good quality point spread function is obtained.

\section{21~cm observables: power spectra and images}
\label{sec:observables}

The ultimate goal of 21~cm observations is to image the spatial distribution of the 21~cm signal as a function of redshift, also known as {\it 21~cm tomography}. Given the curent theoretical predictions, such observations need to achieve mK sensitivity on a few arcminute angular scales (see Chapter~1, 2 and 3 in this book). Most of the current arrays, however, only have the sensitivity to perform a statistical detection of the 21~cm signal, i.e. to measure its power spectrum. Given an intensity field $T$, function of the three dimensional spatial coordinate $\bf x$, its power spectrum $P(k)$ is defined as:
\begin{equation}
\langle {\tilde T}^* ({\bf k}) {\tilde T}({\bf k'}) \rangle = (2 \pi)^3 P(k) \delta^3({\bf k} - {\bf k}')
\label{sec_observables_eq1}
\end{equation}
where $\langle \rangle$ indicates the ensamble average, $\bf k$ is the Fourier conjugate of $\bf x$, tilde the Fourier transform, $^*$ the conjugate operator, $k$ the magnitude of the $\bf k$ vector and $\delta$ the Dirac delta function. In 21~cm observations, power spectra can be computed from interferometric image cubes after deconvolution of the dirty image ${\bar I}_D (l,m,\nu)$ from the point spread function (e.g., \cite{pen09}, \cite{harker10}, \cite{beardsley16}, \cite{patil17}). Alternatively, the 21~cm power spectrum can be estimated directly from the interferometric visibilities. Equation~\ref{eq:4} already shows that the interferometer is a ``natural" spatial power spectrum instrument (e.g., \cite{white99} and Figure~\ref{fig:fig1c}, \ref{fig:fig1d}). Visibilities can be further Fourier transformed along the frequency axis (the so-called {\it delay transform}, \cite{parsons12a}): 
\begin{equation}
{\tilde V}_{ij} (u,v, \tau) = \int_B V (u,v, \nu) \, e^{-2 \pi i \nu \tau} d \nu
\label{sec_observables_eq2}
\end{equation}
where $B$ is the observing bandwidth and the delay $\tau$ is the Fourier conjugate of $\nu$\footnote{The delay variable here is almost equivalent to the geometrical delay and that is why I used the same symbol (see \cite{parsons12a} for details).}. The delay transform is therefore proportional to the three dimensional power spectrum (\cite{parsons12b}):
\begin{equation}
P(k) \propto {\tilde V}_{ij} (|{\bf b}|, \tau),
\label{sec_observables_eq3}
\end{equation}
where the proportionality constant transforms the visibility units into power units (\cite{parsons12b}). The observer units $({\bf b},\tau)$ map directly in $k$ modes perpendicular and parallel to the line of sight (e.g., \cite{morales04}):
\begin{equation}
k_\perp = \frac{2 \pi |{\bf b}|}{D_c} = \frac{2 \pi \sqrt{u^2 + v^2}}{D_c}, \,\,\,\,\,\,\,    k_\parallel = \frac{2 \pi f_{21} H_0 E(z)}{c \, (1 + z)^2} \, \tau,
\label{sec_observables_eq4}
\end{equation}
where $D_c$ is the transverse comoving distance, $f_{21} = 1421$~MHz, $H_0$ is the Hubble constant and $E(z) = \sqrt{\Omega)m (1+z)^3 + \Omega_k (1+z)^2 + \Omega_\Lambda}$. Due to the dependence of the baseline length upon frequency, equation~\ref{sec_observables_eq3} is only valid for short baselines, typically shorter than a few hundresd meters, for which the baseline length can be considered constant across the bandwidth and lines of constant $k_\parallel$ are essentially orthogonal to the $k_\perp$ axis (\cite{parsons12b}).

\begin{figure}[]
\begin{center}
\includegraphics[width=1.\textwidth]{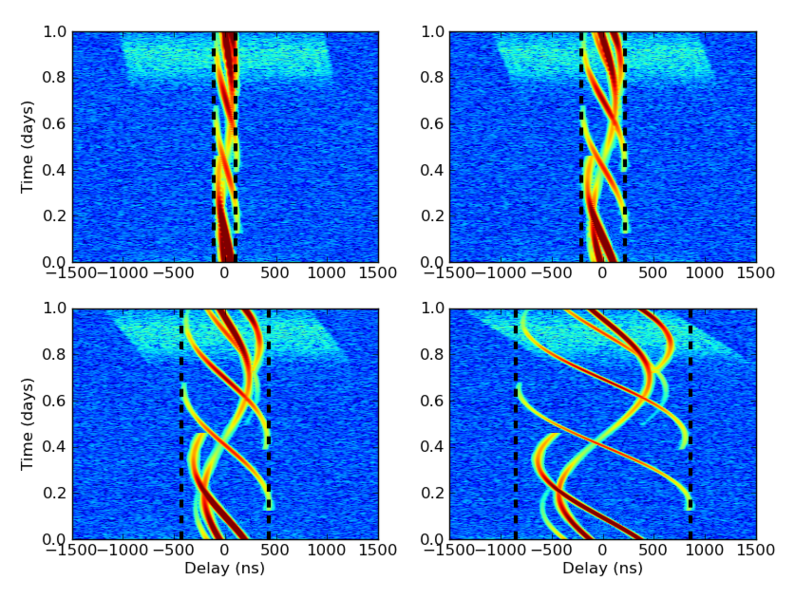}
\end{center}
\caption{Amplitude of delay transformed visibilities as a function of time and delay for a 32 (top let), 64 (top right), 128 (bottom left) and 256~m (bottom right) baseline respectively (from \cite{parsons12a}). A number of smooth spectrum point sources are simulated as foregrounds and their tracks are clearly bound within the horizon limit (black dashed line). The cyan emission is a fiducial 21~cm model that has power up to high delays regardless of the baseline length. The 21~cm signal is, in principle, directly detectable outside the horizon limit (EoR window) without foreground contamination.}
\label{fig:fig2}
\end{figure}

Equation~\ref{sec_observables_eq2} does not only provide a link between visibilities and three dimensional power spectra, but also introduces the concept of ``horizon limit", i.e. the maximum physical delay allowed $\tau_{\rm max} = \frac{|{\bf b}|}{c}$, where $c$ is the speed of light. The most relevant implication of the existence of a horizon limit is the definition of a region in the two dimensional $(k_\parallel,k_\perp)$ power spectrum space where smooth-spectrum foregrounds are confined, leaving the remaining area  uncontaminated in order to measure the 21~cm signal (the so-called ``Epoch of Reionization (EoR) window", Figure~\ref{fig:fig2}). Foregrounds can therefore be ``avoided" with no requirements for subtraction (e.g., \cite{morales12}, \cite{vedantham12}, \cite{pober13}, \cite{thyagarajan13}; see also Chapter~6 in this book).
The choice of a foreground avoidance strategy versus subtraction plays an important role in planning an experiment, its related observing strategy and the array calibration strategy.

The requirements for image tomography are the same as for high brightness sensitivity observations of diffuse emission like the Cosmic Microwave Background (e.g., \cite{halverson02}, \cite{dickinson04}, \cite{readhead04}). The 21~cm spatial distribution throughout cosmic reionization has structures on 5-10 arcminutes up to degree scales (e.g., \cite{majumdar12}, \cite{datta12}, \cite{mellema13}, \cite{kakiichi17}). In order to image 21~cm fluctuations, a maximum baseline of the order of a few km is required to obtain a resolution of a few arcminutes in the $100-200$~MHz range, together with filled $uv$ plane in order to accurately reconstruct their complex spatial structure. A filled $uv$ plane also leads to a point spread function with very low sidelobes, making the deconvolution process easier (see the bottom right panel of Figure~\ref{fig:fig1d} for an example of densely sampled $uv$ plane that leads to a good quality point spread function). The most stringent requirements for image tomography remain the accurate foreground separation and, as I will review in the next section, the related instrumental calibration.

\section{Interferometric calibration and 21~cm observations}
\label{sec:calibration}

Celestial radio signals always experience a corruption when observed with an interferometric array, due to the non-ideal instrumental response that is corrected in post processing in a process that is known as interferometric calibration. Calibration relies on the definition of a data model where the corruptions are described by antenna based quantities known as Jones matrices. Such data model is known as the interferometric measurement equation (\cite{hamaker96},\cite{smirnov11},\cite{smirnov11b},\cite{smirnov11c}).

If antenna $1$ and antenna $2$ measure two orthogonal, linear polarizations $x$ and $y$, the cross-polarization visibility products can be grouped in a $2 \times 2$ complex matrix ${\bf V}$:
\begin{equation}
    {\bf V}_{12} (u,v,\nu) \equiv 
    \left[ 
    \begin{array}{cc}
    V_{12,xx} (u,v,\nu) & V_{12,xy} (u,v,\nu) \\
    V_{12,yx} (u,v,\nu) & V_{12,yy} (u,v,\nu) \\
    \end{array}
    \right].   
\label{eq:sec:1}
\end{equation} 
The sky brightness distribution $I$ can also be written as a $2 \times 2$ matrix ${\bf B}$ using the Stokes parameters as a polarization basis:
\begin{equation}
    {\bf B}_I (l,m,\nu) \equiv 
    \frac{1}{2} \left[
    \begin{array}{cc}
    I (l,m,\nu) + Q (l,m,\nu) & U (l,m,\nu) + iV (l,m,\nu) \\
    U (l,m,\nu) - iV (l,m,\nu) & I (l,m,\nu) - Q (l,m,\nu) \\
    \end{array}
    \right].   
\label{eq:sec:2}
\end{equation} 
At this point, equation~\ref{eq:3} can be written by including the corruptions represented by the complex Jones matrices ${\bf J}$ (\cite{hamaker96},\cite{smirnov11}):
\begin{equation}
{\bf V}_{12} (u,v,\nu) = {\bf J}_1 \left( \int_\Omega {\bf B}_I (l, m, \nu) \, e^{-2 \pi i (ul + vm)} dl \, dm  \right) {\bf J}_2^H,
\label{eq:sec:3}
\end{equation}
where $H$ is the Hermitian operator.
 
Equation~\ref{eq:sec:3} is known as the {\it measurement equation} and is the core of interferometric calibration. For an array with $N$ antennas, equation~\ref{eq:sec:3} can be written for each of the $\frac{N (N - 1)}{2}$ visibilities forming an overdetermined system of equations. The development of  algorithms to solve the calibration system of equations is a very active research line (\cite{mitchell08}, \cite{kazemi11}, \cite{tasse14}, \cite{yatawatta15}, \cite{smirnov15}) although beyond the scope of this chapter and we mention it here for completeness.

The solution of the measurement equation requires some knowledge of the sky brightness distribution ${\bf B}_I$, in other words, a {\it sky model}. Traditionally this is achieved by observing a calibration source, i.e. a bright, unresolved point source with known spectral and polarization properties. Calibration solutions are then applied to the observed field that is then used to improve the sky model ${\bf B}_I$ which, in turn, leads to more accurate calibration solutions ${\bf J}$. This loop is traditionally called selfcalibration (\cite{cornwell81}, \cite{pearson84}) and can lead to a highly accurate calibration (e.g., \cite{bernardi10}, \cite{smirnov11b}).

The advantage of the measurement equation formalism is that it can factorize different physical terms into different matrices. For example, the frequency response of the telescope electronics and its time variations essentially affects only the two polarization responses and are modeled with a diagonal Jones matrix ${\bf B}$:
\begin{equation}
    {\bf B} (t,\nu) \equiv 
    \left[
    \begin{array}{cc}
    b_x (t,\nu) 	& 	0 	\\
    0 		& b_y (t,\nu) 	\\
    \end{array}
    \right],   
\label{eq:sec:4}
\end{equation} 
where we made it explicit that ${\bf B}$ can vary with time and frequency. The undesired instrumental leakage between the two orthogonal polarizations can be written as a ${\bf D}$ Jones matrix of the form:
\begin{equation}
    {\bf D} (t,\nu) \equiv 
    \left[
    \begin{array}{cc}
    1	 		& d_x (t,\nu)	\\
    -d_y (t,\nu)	& 1 	\\
    \end{array}
    \right],   
\label{eq:sec:5}
\end{equation} 
and the measurement equation can be written as:
\begin{equation}
{\bf V}_{12} (u,v,\nu) = {\bf B}_1 \, {\bf D}_1 \left( \int_\Omega {\bf B}_I (l, m, \nu) \, e^{-2 \pi i (ul + vm)} dl \, dm  \right) {\bf D}_2^H  \, {\bf B}_2^H.
\label{eq:sec:6}
\end{equation}
We note that, in principle, the primary beam response should appear as an additional $2 \times 2$ Jones matrix before the ${\bf D}$ matrix. I have ignored it for now, although I will discuss it late rin this section.

Retaining only the first order terms, equation~\ref{eq:sec:6} can be written as (\cite{sault96}):
\begin{eqnarray}
V_{12,xx} (u,v,\nu) & = & b_{1,x} \, b_{2,x}^* [V_I (u,v,\nu) - V_Q (u,v,\nu)]\\
V_{12,xy} (u,v,\nu) & = & b_{1,x} \, b_{2,y}^* [(d_{1,x} - d_{2,y}^*) V_I (u,v,\nu) + V_U (u,v,\nu) + iV_V (u,v,\nu)]	\\
V_{12,yx} (u,v,\nu) & = & b_{1,y} \, b_{2,x}^* [(d_{2,x} - d_{1,y}^*) V_I (u,v,\nu) + V_U (u,v,\nu) - iV_V (u,v,\nu)]	\\
V_{12,yy} (u,v,\nu) & = & b_{1,y} \, b_{2,y}^* [V_I (u,v,\nu) - V_Q (u,v,\nu)],
\label{eq:sec:7}
\end{eqnarray}
where I dropped the explicit dependence on time and wavelength from the gain terms for notation clarity, and where $V_{i=I, Q, U, V}$ are the Fourier transforms of the elements of the sky brightness matrix ${\bf B}_I$.

This form of the measurement equation offers an intuitive understanding as to why calibration is of paramount importance in 21~cm observations. The observed visibilities are essentially a measurement of foreground emission and, in the ideal case, their amplitudes would vary smoothly with frequency, and foregrounds could either be avoided or subtracted. However, the instrumental response inevitably corrupts this smoothness in several ways: because the telescope primary beam is not sufficiently smooth in frequency, because of the electronic response or because of reflections along the signal path. Although calibration will correct for these effects and restore the intrinsic foreground frequency smoothness, calibration errors (i.e., deviations from the true ${\bf B}$ and ${\bf D}$ solutions) will still corrupt the foreground spectra. In practice, calibration errors result in foreground power leaking out of the horizon limit and jeopardizing (part of) the EoR window. The corruption of foreground spectra will limit the accuracy of any subtraction method (see discussion in Chapter~6 in this book). {\it The effectiveness of foreground separation, proven in ideal cases, depends significantly on the accuracy of interferometric calibration.}

The form of the measurement equation written in equations~\ref{eq:sec:6} and \ref{eq:sec:7} is often referred to as a {\it direction independent} calibration as it implicitly assumes that a single Jones matrix is sufficient to describe corruptions across the whole sky area of interest. This assumption is often invalid at low frequencies, mostly because of the changing primary beam response over a wide field of view, frequency, and over the course of the observation, and the position and time dependent corruptions introduced by the Earth ionosphere. In this case the measurement equation becomes {\it direction dependent}, i.e. a different Jones matrix is written and solved for a certain number of directions in the sky:
\begin{equation}
{\bf V}_{12} (u,v,\nu) = \sum_s \left[ {\bf B}_{1,s} \left( \int_\Omega {\bf B}_{I,s} (l, m, \nu) \, e^{-2 \pi i (ul + vm)} dl \, dm  \right) {\bf B}_{2,s}^H \right],
\label{eq:sec:7_added}
\end{equation}
where the sum is over the number of directions $s$. We note that we have used the ${\bf B}$ matrix for pedagogical purposes here, regardless of the physical origin of the direction dependent effect. Direction dependent effects also impact foreground separation, in a similar way as the direction independent effects.
 
Accurate direction independent and dependent calibration of 21~cm observations is at the forefront of current research and can be grouped in a few main topics:
\begin{itemize}
\item {\it sky models}. Ideally, the sky brightness model matrix ${\bf B}_I$ (equation~\ref{eq:sec:6} and \ref{eq:sec:7}) would include the whole sky emission. This is pratically impossible as part of the sky signal is the unknown of interest (the 21~cm signal) and the detailed properties of the foreground sky are not known sufficiently well. \begin{figure}[]
\begin{center}
\includegraphics[width=1.\textwidth]{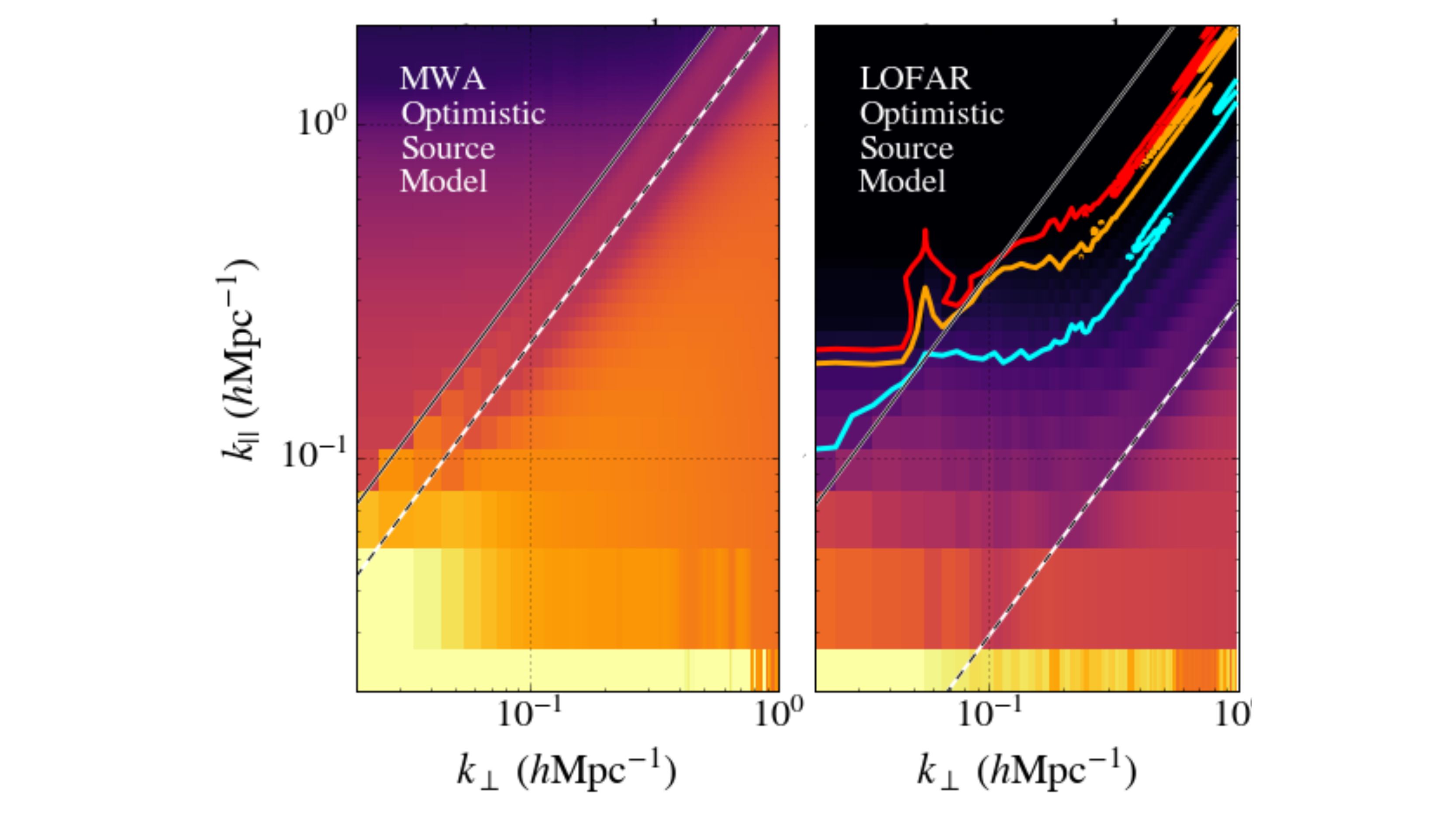}
\end{center}
\caption{Example of power spectrum bias introduced by calibration errors due to an incomplete sky model for the Murchison Widefield Array (MWA, left) and the Low Frequency Array (LOFAR, right) cases respectively (adapted from \cite{ewall-wice17}). Power spectra are shown in their two dimensional $(k_\perp,k_\parallel)$ form in order to display the foreground dominated region below the horizon limit (grey solid line). Cyan, orange and red lines are the locii where a fiducial 21~cm model power spectrum is one, five and ten times higher than the bias level. In an ideal case with perfectly smooth foregrounds and no calibration errors, the 21~cm power spectrum should be detectable just outside the horizon limit. The errors introduced by an incomplete sky model leak foreground power in the EoR window at a level that may completely prevent a detection in the MWA case.}
\label{fig:fig_added}
\end{figure}
Sky models are normally constituted of a catalogue of compact sources of known (or measured) properties, often covering an area significantly larger than the telescope field of view (e.g., \cite{yatawatta13}, \cite{pober16}). Nevertheless, sky models remain essentially always incomplete at some level, as source catalogues are limited in depth, source characterization and - often - sky coverage. \cite{grobler14}, \cite{wijnholds16} and \cite{grobler16} show that incomplete catalogues used as sky models bias the calibration and eventually lead to artifacts in the form of ghost-like sources in interferometric images, most of the times fainter than the image noise level. The ghost pattern is stronger for regularly spaced arrays and if the sky model is less complete. In terms of power spectrum, \cite{ewall-wice17} and \cite{barry16} show that the calibration bias introduced by incomplete sky models leads to an overall leakage of foreground power in the EoR window (Figure~\ref{fig:fig_added}). A similar foreground leakage may occur because of the finite angular resolution of interferometric observations: for example, two sources whose size is respectively one third and one tenth of the instrument angular resolution will both be modeled as point like even if the first source is only barely unresolved. This biased catalogue would again lead to a leakage of foreground power in the EoR window (\cite{procopio17}). In this case, the bias can be mitigated by obtaining a sky model with an angular resolution that is much higher than the scales at which the 21~cm signal is expected (\cite{procopio17}).

Sky models that include only compact sources are not adequate for baselines shorter than a few tens of meters as they are sensitive to Galactic diffuse emission, which contributes to most of the power on angular scales $\theta > 10-20$~arcmin (e.g., \cite{bernardi09}, \cite{choudhuri17}). Excluding short baselines from the calibration solutions prevents the problem of modeling diffuse emission, but can bias the solutions (\cite{patil16}) if the system of calibration equation is not properly constrained, e.g. via regularization (\cite{sardabaradi19}).

In summary, different analysis approaches provide evidence that {\it imperfect sky models (either because of missing catalogue sources, mis-estimating source properties or missing diffuse emission) are a source of calibration bias that has general effect to corrupt the foreground properties, leaking their power well beyond the ideal horizon limit} and requiring additional modeling and subtraction. For this reason, significant efforts are currently ongoing in order to improve sky models via wider and deeper low frequency surveys (e.g., \cite{hurley-walker17}, \cite{intema17}, \cite{shimwell19}), more accurate low frequency catalogues (\cite{carroll16}) and even better observations of Galactic diffuse emission (\cite{zheng17}, \cite{dowell17}); 

\item {\it instrument/primary beam models}. A complete knowledge of a sky model may not be, by itself, sufficient for an accurate calibration of 21~cm observations as the brightness matrix ${\bf B}_I$ is multiplied by the antenna primary beam (equation~ \ref{eq:4} and \ref{eq:sec:3}) and the measurement of an intrinsic sky model requires the separation from the primary beam effect. 

Unlike steerable dishes, most 21~cm interferometers are constituted of dipoles fixed on the ground, in some cases clustered together to form larger stations whose beams that can be digitally pointed to a sky direction by introducing different delays to the dipoles (e.g., like the MWA and LOFAR arrays). As station beams are formed in order to track a source on the sky, the station projected area changes with time and the shape of the primay beam changes noticeably (Figure~\ref{fig:fig3}). This is a typical direction dependent effect that can be casted in the measurement equation as
\begin{equation}
{\bf V}_{12} (u,v,\nu) = \int_\Omega {\bf E}_1 (t, l, m, \nu) {\bf B}_{I,s} (l, m, \nu) \, e^{-2 \pi i (ul + vm)} \, {\bf E}_2^H (t, l, m, \nu)  dl \, dm,
\label{eq:sec:7_added_2}
\end{equation}
were ${\bf E} (t, l, m, \nu)$ is the Jones matrix describing the primary beam which, in the simplest cases, is a diagonal matrix:
\begin{equation}
    {\bf E} (t, l, m, \nu) \equiv 
    \left[
    \begin{array}{cc}
    e_x (t, l, m, \nu) 	& 	0 	\\
    0 		& e_y (t, l, m, \nu) 	\\
    \end{array}
    \right].
\label{eq:sec:4}
\end{equation} 
We note that we have written the explicit dependence on the time due to change in projected area for dipole stations and that the direction dependence of the ${\bf E}$ is encoded in its $(l, m)$ dependence.

Time and frequency variable primary beams lead to apparent time variable sky models with variations that are larger away from the pointing direction due to the greater changes in the sidelobe pattern. For examples, sky sources that are well within the main lobe of the primary beam in Figure~\ref{fig:fig3} will experience relatively negligible variations throughout an observation, the opposite will occur to sources located well outside the main lobe as they run through primary beam sidelobes.

Primary beams are also frequency variable and, to first order, their size scales with the observing wavelength (equation~\ref{eq:2}), i.e. rather smoothly. However, in the sidelobe region, variations become rather abrupt as the source can be located on a sidelobe peak at a certain frequency and in the sidelobe null at another frequency. As a final remark, stations that include several dipoles are not perfectly equal to each other, due to manufacturing reasons or mutual coupling between their elements (e.g., \cite{sokolowski17}), leading to  ${\bf E}_1 \neq {\bf E}_2$. As primary beams are different, even visibilities for baselines that have the same length and orientation will be different - rather than identical, as expected. The left panel of Figure~\ref{fig:fig3} shows an example of how much primary beams vary for different stations due to mutual coupling interactions: variations in the sidelobe region can be as large as $\sim 30\%$.

If not accurately modeled and taken into account, primary beam effects can bias the calibration solution and, again, corrupt the foreground frequency smoothness.  \cite{bhatnagar08}, \cite{bernardi11}, \cite{sullivan12} and \cite{tasse13} have developed methods to incorporate time and frequency variable primary beams in interferometric images, however, the accuracy of the correction is limited by the accuracy of the primary beam model. Increasing effort is therefore being placed in precise modeling and measurements of primary beams (e.g., \cite{pupillo15}, \cite{trott16}, \cite{deleraacedo17}, \cite{trott17a}, \cite{jacobs17}, \cite{deleraacedo18});
\begin{figure}[]
\begin{center}
\includegraphics[width=1.\textwidth]{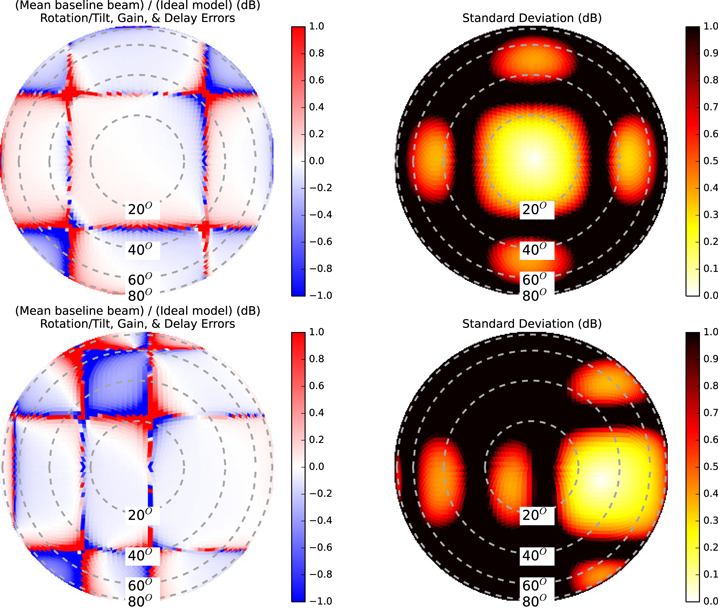}
\end{center}
\caption{Example of primary beam variations as an MWA station points at zenith (top right) and $\sim 30^\circ$ away from zenith (bottom right) at 150~MHz. The left column shows the fractional variation of individual station beam models, with respect to the nominal primary beam (right column, from \cite{neben16}). It is visibile how different the sidelobe pattern is when pointing towards two different directions. The $\sim 10\%$ magnitude of the first lobe and the large null regions around the sidelobes should also be noticed. The specific pattern is due to the regular shape of the MWA station, where 16 dipoles are arranged in a square $4 \times 4$ grid.}
\label{fig:fig3}
\end{figure}

\item {\it polarization leakage calibration}. Equation~\ref{eq:sec:3} and \ref{eq:sec:7} show that, even if the 21~cm signal is unpolarized, care needs to be taken against the contamination from polarized foreground emission. Most point sources are unpolarized below 200~MHz (\cite{bernardi13}, \cite{lenc16}, \cite{vaneck18}), therefore the assumption of an unpolarized sky model is well justified. However, calibration errors (in the ${\bf B}$ matrix) would lead  to a relative miscalibration of the $xx$ and $yy$ polarizations and, in turn, to leakage of polarized emission into total intensity. This effect may be particularly strong on short baselines (e.g., shorter than a $\sim 1$~km), where polarized foregrounds are brighter (\cite{bernardi09}, \cite{iacobelli13}, \cite{jelic15}, \cite{lenc16}). Polarized foregrounds that are Faraday rotated by the interstellar medium and leak to total intensity are a severe contamination to the 21~cm signal: they have a characteristic frequency dependence similar to the 21~cm signal therefore have power across the whole EoR window and cannot be subtracted using standard methods (e.g., \cite{jelic10}, \cite{moore13}, \cite{nunhokee17}).  

Even if calibration errors are negligible, low frequency antennas have a non negligible polarized response across their wide field of view, i.e. the primary beam Jones matrix ${\bf E}$ is no longer diagonal. The measurement equation with a full polarized primary beam response can be written as (\cite{nunhokee17}):
\begin{eqnarray}
    \left[ 
    \begin{array}{cc}
    V_{12,I} (u,v,\nu)\\
    V_{12,Q} (u,v,\nu)\\
    V_{12,U} (u,v,\nu)\\
    V_{12,V} (u,v,\nu)
    \end{array}
    \right] & = & \int_\Omega{ {\bf S}^{-1} [ {\bf E}_1 \otimes {\bf E}_2^H ] {\bf S} 
    \left[ 
    \begin{array}{cc}
    I (l,m,\nu)\\
    Q (l,m,\nu)\\
    U (l,m,\nu)\\
    V (l,m,\nu)
    \end{array}
    \right]
     \, e^{-2 \pi i (ul + vm)} dl \, dm } = \nonumber \\
     & = & \int_\Omega{ {\bf A} (l,m,\nu) 
    \left[ 
    \begin{array}{cc}
    I (l,m,\nu)\\
    Q (l,m,\nu)\\
    U (l,m,\nu)\\
    V (l,m,\nu)
    \end{array}
    \right]
     \, e^{-2 \pi i (ul + vm)} dl \, dm },
\label{eq:sec:8}
\end{eqnarray} 
where ${\bf S}$ is the matrix that relates the intrisic Stokes parameters to the observer $x-y$ frame (\cite{hamaker96}) and $\otimes$ is the outer product. Visibilities are written as a four-element vector as this form shows that the ${\bf A}$ matrix maps the intrinsic (unprimed) Stokes parameters into the observed (primed) ones: 
\begin{eqnarray}
    \left(
    \begin{array}{cccc}
    I' \leftarrow I & I' \leftarrow Q & I' \leftarrow U & I' \leftarrow V \\
    Q' \leftarrow I & Q' \leftarrow Q & Q' \leftarrow U & Q' \leftarrow V \\
    U' \leftarrow I & U' \leftarrow Q & U' \leftarrow U & U' \leftarrow V \\
    V' \leftarrow I & V' \leftarrow Q & V' \leftarrow U & V' \leftarrow V \\
    \end{array}
    \right).
\label{eq:sec:9}
\end{eqnarray} 
An example of ${\bf A}$ matrix is shown in Figure~\ref{fig:fig4}. The first row of the matrix shows how the four intrisinc Stokes parameters contribute to the observed total intensity and, therefore, how polarized foregrounds leak into the 21~cm signal even in absence of any calibration errors: the magnitude of the contaminating Stokes $Q$ and $U$ foregrounds increases away from the pointing direction. Wide-field polarization is another textbook example of direction dependent calibration problem.

Calibration of polarization leakage remains a challenging task. Instruments with narrow fields of view are less prone to polarization leakage (\cite{asad15}, \cite{asad16}, \cite{asad18}). Another way of mitigating polarization leakage is extend the sky model to include polarization (e.g. \cite{geil11}), although modeling the diffuse Galactic foreground - the brightest component - is not straightforward and requires accurate imaging. \cite{nunhokee17} show, however, that the magnitude of the Galactic polarization leakage may be below the 21~cm signal at high $k_\parallel$ values ($k_\parallel > 0.3$~Mpc$^{-1}$) and, potentially, an avoidance strategy is not completely excluded. A more extensive characterization of the polarized foreground properties is needed in order to generalize their results.
\begin{figure}[]
\begin{center}
\includegraphics[width=1.\textwidth]{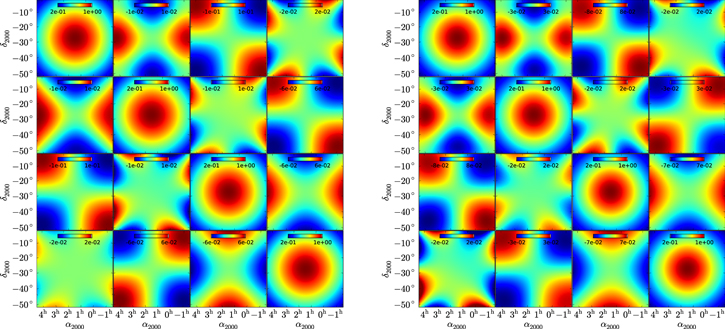}
\end{center}
\caption{Examples of ${\bf A}$ matrices that model the dipole of the Precision Array to Probe the Epoch of Reionization (PAPER, \cite{parsons10}) at 130 (left) and 150~MHz (right) respectively. They map the intrinsic Stokes parameters into observed ones: the diagonal terms represent the standard primary beam patterns, whereas the off-diagonal terms are the leakage terms. The second, the third and the fourth element of the first row show how Stokes parameters $Q$, $U$ and $V$ respectively contaminate the total intensity signal (from \cite{nunhokee17}).}
\label{fig:fig4}
\end{figure}

\begin{figure}[]
\begin{center}
\includegraphics[width=1.\textwidth]{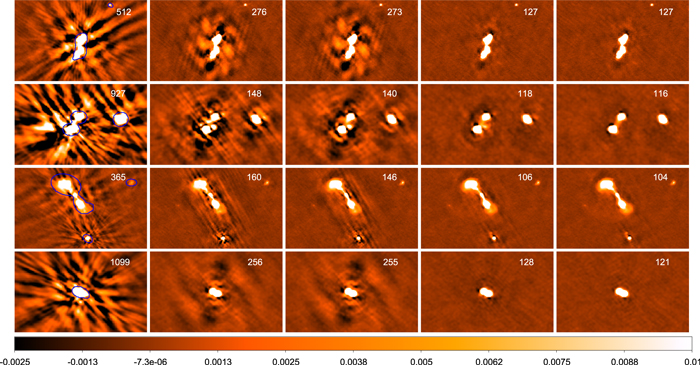}
\end{center}
\caption{Calibration of ionospheric effects in LOFAR observations using a faceting algorithm (from \cite{vanweeren16}). The image resolution is $8'' \times 6''.5$ averaged over the $120-180$~MHz bandwidth. The left column shows zoom in images around sources without direction dependent calibration which is, in turn, applied incrementally towards the right panels. For each source, a sky model and a direction dependent Jones scalar $Z$ is improved at each iteration until an artefact-free image is obtained (right column). Solutions were computed every 10~s. An additional amplitude calibration to account for primary beam variations was determined on scales of 10~minutes. The colour scale is in units of Jy~beam$^{-1}$.}
\label{fig:fig5}
\end{figure}
\item {\it ionospheric distortions}. The ionosphere is the partially ionized layer situated between $\sim 50$ and 1000~km above the surface of the Earth, whose electron density changes with time and position. At low frequencies the ionosphere is no longer transparent to radio waves and, to first order, it  delays the wave propagation by an amount proportional to the integral of the electron density along the line of sight (e.g. \cite{TMS}, \cite{intema09}):  
\begin{equation}
\phi (t, \nu) \propto \frac{1}{\nu} \int n_e (t) dl,
\label{eq:sec:11}
\end{equation}
where $\phi$ is the extra delay, $n_e$ the electron density and the integral is the total electron content (TEC) along the line of sight. When the delay is different for two different antennas, visibilities measure an additional, time variable delay. In the measurement equation formalism, ionospheric delays can be modeled by a scalar term $Z \propto e^{i\phi (t,\nu)}$, however, ionospheric effects are another texbook example of direction dependent calibration as the $Z$ is different for different directions. Given the size $S$ of a characteristic ionospheric patch where the TEC is constant, direction dependent effects occur when either the field of view is much larger than $S$ or the baseline separation is much larger than $S$, i.e. different antennas ``see through" different TEC values (see \cite{intema09} for an extensive discussion on the different ionospheric regimes). In this case, the measurement equation takes a form similar to equation~\ref{eq:sec:7_added}: 
\begin{equation}
{\bf V}_{12} (u,v,t,\nu) = \sum_s Z_{1,s} (t, \nu) \left ( \int_\Omega {\bf B}_{I,s} (l, m, \nu) \, e^{-2 \pi i (ul + vm)} \,  dl \, dm \right ) Z_{2,s}^H (t, \nu),
\label{eq:sec:12}
\end{equation}
leading to images where sources are convolved with a position and time dependent point spread function. An example of this effect is shown in Figure~\ref{fig:fig5}: the column on the left shows sources after the standard selfcalibration, still surrounded by artifacts due to the ionosphere; moving towards the right,  iterative direction dependent corrections lead to virtually artefact-free images on the right column (see \cite{vanweeren16} for further details). 

\cite{trott17b} analyzed the effects of ionospheric perturbations on MWA observations, whose maximum baseline is a factor of $\sim 30$ shorter than the LOFAR example displayed in Figure~\ref{fig:fig5}, but with a field of view $\sim 4$ times larger. They found that direction dependent ionospheric distortions can affect the sky coherence up to degree-scales (i.e. scales relevant for 21~cm observations), however, due to the relatively short baselines, these effects occur only in 8\% of the observations and it is relatively straightforward to monitor the ionospheric activity and exclude the most affected observations.

An extensive modeling of the impact of ionospheric errors on the two dimensional $(k_\perp,k_\parallel)$ power spectrum has been carried out by \cite{vedantham16}. They found that most of the residual effects due to the ionosphere on baselines shorter than a few km are confined within the horizon limit, therefore not impacting foreground avoidance. Moreover, the frequency coherence of the ionospheric residual errors is such that they will likely be removed by foreground subtraction algorithms.

Current investigations seem therefore to suggest that ionospheric effects are not going to be a show stopper for both 21~cm power spectrum observations and, likely, 21~cm tomography. 
\end{itemize}

\subsection{Redundant calibration}

An interferometric array where most of the baselines have the same length and orientation is called {\it redundant}, as these baselines measure the same Fourier mode of the sky brightness distribution. Redundant array configurations are often not appealing as they have poor imaging performances because they do not measure sufficient Fourier modes to reconstruct accurate sky images. However, a maximally redundant array where the antennas are laid out in a regularly spaced square grid offers the maximum power spectrum sensitivity on the $k_\perp$ modes corresponding to the most numerous baselines. This criterium has inspired the highly redundant layouts of the MIT Epoch of Reionization experiment (\cite{zheng14}), PAPER (\cite{parsons12b}) and partly driven the updated MWA (\cite{wayth18}).

One of the advantages of a redundant array is that it enables a different calibration strategy, i.e. {\it redundant calibration}. In redundant calibration the form of the measurement equation does not change and can be written, for a single polarization, like equation~\ref{eq:sec:7}:
\begin{equation}
V_{12,xx} (u,v,\nu) = b_{1,x} \, b_{2,x}^* \, Y_{12,xx} (u,v,\nu),
\label{eq:sec:13}
\end{equation}
with the difference now that the model visibility $Y$ is not tied to a sky model, but it is solved for, simply assuming that it is the same for each group of redundant baselines ({\cite{wieringa92}, \cite{liu10}). In other words, redundant calibration is independent on the sky model and, therefore, bypasses entirely the biases related to sky model incompleteness described in Section~\ref{sec:calibration}. However, as redundant calibration is not tied to any physical (i.e. sky-based) spatial or spectral model, its solutions have degeneracies that need to be solved for by using a sky model (e.g., \cite{zheng14}, \cite{byrne19}). In particular, spectral calibration, which is critical for foreground separation, cannot currently be obtained using redundant calibration and requires a sky-based calibration. \cite{byrne19} suggest that sky model incompleteness can bias this calibration step, in a way similar to what happens with a traditional calibration scheme. Moreover, as redundant calibration is agnostic of the polarization state of the sky brightness distribution, mitigation of polarization leakage remains an open question in the framework of redundant calibration (\cite{dillon18}). 

Finally, redundant calibration is prone to effects that break the assumption of redundancy, the most common being errors in the antenna positions and different antenna primary beams. Antenna position errors can be reduced to have a negligible impact on redundant calibration \cite{joseph18}. The effect of primary beam variations amongst the different antennas on redundant calibration is likely more severe, although new calibration schemes are being developed to mitigate it (\cite{orosz19}).

\section{Array design and observing strategies}
\label{sec:design}

\begin{figure}[]
\begin{center}
\includegraphics[width=1.\textwidth]{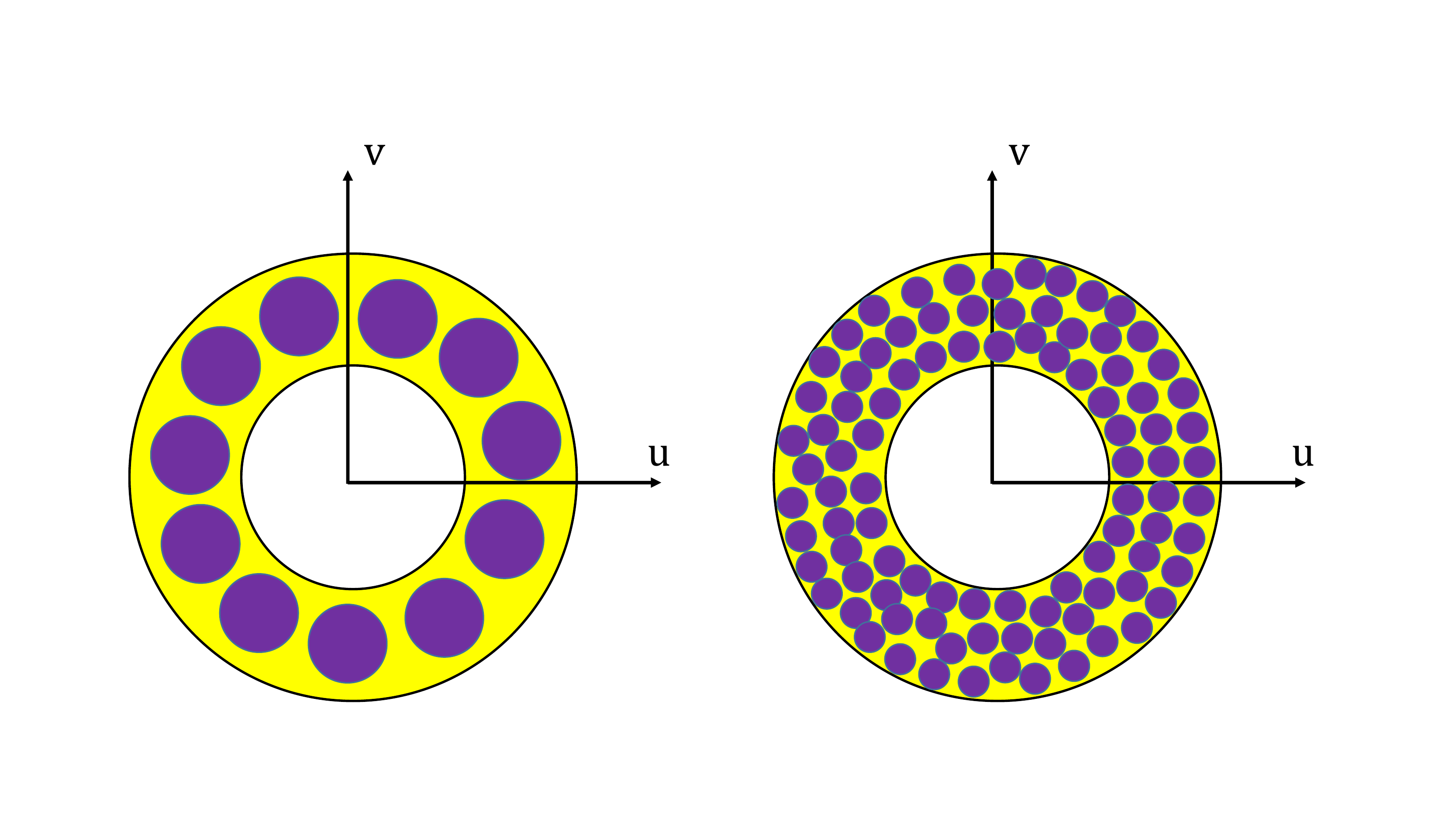}
\end{center}
\caption{Cartoon illustration of the $uv$ footprint due to the primary beam. The purple circles are the $uv$ footprints for a large (left panel) and small (right panel) station respectively. The minimum and maximum baselines is the same for both cases, in order to sample the same $uv$ annulus (yellow area).}
\label{fig:fig6}
\end{figure}
I will conclude this chapter by discussing how the various interferometric effects discussed so far impact the choice of array designs and the consequent observing strategies. \cite{morales05} and \cite{parsons12b}, for example, investigate how instrumental choices like the array layout, the antenna size and the bandwidth (do not) affect measurements of the 21~cm power spectrum spectrum. Here I would rather emphasize the interdependence between instrumental choices, calibration and foreground separation strategies. If the total collecting area is kept fixed, there are two main elements that impact calibration and foreground separation strategies:
\begin{itemize}
\item {\it station size}. The choice of the station size determines the minimum $k_\perp$ value accessible and the footprint of each $uv$ measurement. Each visibility is not a single point in the $uv$ plane but has a footprint corresponding to the two dimensional Fourier transform of the primary beam. This can be seen using the convolution theorem to re-write equation~\ref{eq:1}: \begin{equation}
V_{ij} ({\bf b}, \nu) = \tilde{A} ({\bf b}, \nu) \ast \tilde{I} ({\bf b}, \nu).
\end{equation}
Smaller stations have smaller footprints in the $uv$ plane (see Figure~\ref{fig:fig6}) and can, therefore, sample the $uv$ plane more accurately than larger stations. They also allow to probe smaller $k_\perp$ values (as the minimum possible $uv$ length is essentially the station size) for which the avoidance strategy is more effective (see Figure~\ref{fig:fig2}).
If smaller stations are preferred for power spectrum measurements, they are generally more challenging in terms of calibration: they have wider fields of view that require a more accurate sky model for calibration and that suffer from more severe ionospheric distortions and polarization leakage contamination. Given the smaller size, their visibilities have a lower signal-to-noise ratio compared to larger stations, possibly limiting the calibration of high time variable effects. On the other hand, they do not necessarily require to track sources with a high time cadence but can use drift scan strategies (where they are pointed to a fixed direction and the sky drifts overhead) or a mix of drift scan and pointed observations to maximize sensitivity (\cite{trott14}). The advantage of drift scan over pointed observations is that primary beams remain constant in time, avoiding some of the effects described in the Section~\ref{sec:calibration}.

\item {\it array layout}. Beyond the obvious sensitivity requirement that prefers compact arrays due to their better brightness sensitivity, layout choices are also intrinsically related to calibration and foreground separation strategies. A pseudo-random station distribution that leads to a filled $uv$-coverage (between the minimum and the maximum station separation) is highly desirable for imaging, modeling and subtracting foregrounds. It is not a stringent requirement for power spectrum measurement and for the avoidance strategy. It is probably necessary for 21~cm tomography, in order to provide reconstruction of the low-brightness neutral Hydrogen regions.

On the opposite side of the spectrum of choices, redundant arrays are the most sensitive power spectrum machines. They obviously leverage on redundant calibration which is precluded to imaging arrays. Their drawbacks are the poor imaging performances that prevent the accurate foreground modeling and essentially only allow foreground avoidance. For the same reason, if redundant calibration is not sufficient, redundant arrays have limited options to improve calibration by reconstructing the sky brightness sensitivity.   
\end{itemize}

I will use four existing low frequency arrays as examples of the range of cases of interest:
\begin{itemize}

\item {\it Low Frequency Array (LOFAR, \cite{vanhaarlem13})}. LOFAR is an array of 40 stations located in The Netherlands and several remote stations across Europe. 24 stations are located in a 2~km core from the array centre and the remaining stations are distributed in a logarithmic spiral layout up to $\sim 100$~km, providing a very dense $uv$ coverage in a few hours tracked observation (see Chapter~8 in this book for an image of the LOFAR array layout and the other arrays discussed here).

Stations are formed by two types of receptors sensitive to the $30-90$ and $110-200$~MHz range respectively. The $110-200$~MHz stations are the most sensitive to 21~cm observations and we will only consider them in this discussion. They are constituted by 48 clusters of dipoles (each of them being a $4 \times 4$ square grid) arranged in a regular $\sim 30$~m diameter grid, leading a $\sim 4^\circ$ field of view at 150~MHz.

LOFAR is an example of a traditional interferometric array, with excellent point source sensitivity that favours sky-based calibration and a very dense $uv$ coverage for high fidelity imaging. Its large station size has a large $uv$ footprint and a relatively narrow field of view that essentially requires tracking a sky patch. The narrow field of view allows to select sky patches with low foreground (including polarization) contamination and to reject wide field foreground emission. Unwanted sky emssion far from the pointing direction is further suppressed by rotating each station grid with respect to another, while rotating the dipoles back to a common polarization frame: this operation makes the station primary beams all different and their sidelobe patterns, that would otherwise be reinforced by the regular station grid, tend to average out.

The calibration of LOFAR 21~cm observations relies on an accurate sky model where compact sources are modeled using the longest baselines available (i.e. $\sim 100$~km). Direction dependent calibration corrects for ionospheric effects that corrupt visibilities on baselines longer than a few km, and for the effect of variable primary beams on compact sources (\cite{yatawatta13}). The sky model is then subtracted from the visibilities and residual foregrounds are subtracted in the image domain (see details in Chapter~6 in this book). 

The LOFAR design is suited for 21~cm tomography on large angular scales, providing foregrounds are adequately subtracted (\cite{zaroubi12}).

\item {\it Murchison Widefield Array (MWA, \cite{tingay13}, \cite{wayth18})}. The MWA is an array located in Western Australia, operating between 80 and $\sim 200$~MHz. It employs the same LOFAR dipoles, although they are assembled in stations of $4 \times 4$ elements arranged in a regular grid. The station size is therefore $\sim 6$ times smaller compared to LOFAR, with an equivalent increase of the field of view. 
The MWA underwent a recent upgrade to phase II (to distinguish it from the initial deployment, named phase I) and is now constituted of 256 stations (out of which only 128 can be simultaneously correlated) in a hybrid configuration: 128 stations are deployed in a pseudo random configuration out to a $\sim 3$~km baseline (the phase I telescope), 72 stations in two highly redundant hexagons next to the core of the array and 56 stations to extend the maximum baseline up $\sim 5$~km. 

MWA phase II is a fairly versatile instrument: in its compact, redundant configuration, it is optimized for power spectrum observations and can leverage redundant calibration (\cite{li18}); its small stations give a good sampling in the $uv$ plane (right panel case in Figure~\ref{fig:fig6}). In its extended configuration it has an exceptionally good instantaneous $uv$ coverage (due to the high number of stations instantaneously correlated) with low sidelobe levels, which is good for imaging and foreground modeling, and a large field of view which allows to survey the sky very quickly. The wide field of view does not allow to isolate low foreground patches, but it allows to opt for drift scan observations or a mix of drift scan and pointed observations (\cite{trott14}), which have the advantage of more time stable primary beams. Wide field ionospheric effects are somewhat mitigated by the array compactness (\cite{jordan17}).
The MWA can therefore leverage on the strength of both redundant and traditional calibration and can adopt a mixture of foreground subtraction and avoidance strategies.

The MWA approach has, however, limitations too: the regular station grid (without any rotation, unlike LOFAR) generates strong sidelobes (see Figure~\ref{fig:fig3}) which make calibration and foreground separation more challenging; the large field of view requires more comprehensive sky models for calibration and is more susceptible to polarization leakage; the relatively short maximum baseline may be insufficient to derive accurate, high-angular resolution sky models (\cite{procopio17}). 

\item {\it Precision Array to Probe the Epoch of Reionization (PAPER, \cite{parsons10})}. PAPER was an array located in the South Africa, operating in the $100-200$~MHz range and now decommissioned in favour of its successor (the Hydrogen Epoch of Reionization Array, see Chapter~9 in this book). It employed custom designed $\sim 2$~m dipoles that were deployed and re-arranged in several configurations up to a 128~element array. Dipoles were always individually correlated with no clustering into larger stations, implying a nearly all-sky field of view. In order to maximize power spectrum sensitivity, dipoles were always deployed in maximally redundant configuration with very short baselines (up to a maximum of 350~m), enabling the advantages of redundant calibration (\cite{parsons14}, \cite{ali15}, \cite{jacobs15}). In the final 128-element deployment, $\sim 20$ dipoles were placed as outriggers outside the regular grid in order to partially improve the $uv$ coverage for foreground characterization and calibration. 

In some sense, PAPER represents the choice opposite to the LOFAR case: an almost fully redundant array that works using essentially only foreground avoidance and without any spatial characterization of foregrounds for either calibration or subtraction. PAPER is a full drift scan array with primary beams that are fairly stable with time, but also with an all-sky field of view where no selection of low foreground regions is possible, for which polarization leakage and ionospheric effects are the most severe, although the latter are mitigated by the very compact configuration. 

As pointed out earlier in this chapter, a redundat array like PAPER is not suited for 21~cm tomography.
\end{itemize}

\section{Conclusions}
\label{sec:conclusions}

This chapter presented a summary of interferometry and calibration in the light of 21~cm observations. I started from the basics of interferometry to show how they are related to observations of the 21~cm power spectrum and its tomographic images. I reviewed calibration of 21~cm observations, highlighting how foreground separation - the biggest challenge of 21~cm observations - critically depends on various calibration effects (sky models, primary beam modeling and calibration, polarization leakage, the ionosphere). I also attempted to show how the various array designs adopted by current experiments enable different calibration and observational strategies - neither of which is clearly winning, at the present point. The field is rapidly developing and both current and upcoming instruments (see Chapter~9 in this book) will address some of the open questions presented in this chapter.

\section{Acknowledgements}

It is my pleasure to thank C.D. Nunhokee for useful discussions and help with Figure~\ref{fig:fig1c} and \ref{fig:fig1d}.

\bibliographystyle{plain}
\bibliography{Bernardi/References}

%% file: ms.bbl
\begin{thebibliography}{100}

\bibitem{ali15}
Z.~S. {Ali}, A.~R. {Parsons}, H.~{Zheng}, J.~C. {Pober}, A.~{Liu}, J.~E.
  {Aguirre}, R.~F. {Bradley}, G.~{Bernardi}, C.~L. {Carilli}, C.~{Cheng}, D.~R.
  {DeBoer}, M.~R. {Dexter}, J.~{Grobbelaar}, J.~{Horrell}, D.~C. {Jacobs},
  P.~{Klima}, D.~H.~E. {MacMahon}, M.~{Maree}, D.~F. {Moore}, N.~{Razavi},
  I.~I. {Stefan}, W.~P. {Walbrugh}, and A.~{Walker}.
\newblock {PAPER-64 Constraints on Reionization: The 21 cm Power Spectrum at z
  = 8.4}.
\newblock {\em \apj}, 809:61, August 2015.

\bibitem{asad18}
K.~M.~B. {Asad}, L.~V.~E. {Koopmans}, V.~{Jeli{\'c}}, A.~G. {de Bruyn}, V.~N.
  {Pandey}, and B.~K. {Gehlot}.
\newblock {Polarization leakage in epoch of reionization windows - III.
  Wide-field effects of narrow-field arrays}.
\newblock {\em \mnras}, 476:3051--3062, May 2018.

\bibitem{asad16}
K.~M.~B. {Asad}, L.~V.~E. {Koopmans}, V.~{Jeli{\'c}}, A.~{Ghosh}, F.~B.
  {Abdalla}, M.~A. {Brentjens}, A.~G. {de Bruyn}, B.~{Ciardi}, B.~K. {Gehlot},
  I.~T. {Iliev}, M.~{Mevius}, V.~N. {Pandey}, S.~{Yatawatta}, and S.~{Zaroubi}.
\newblock {Polarization leakage in epoch of reionization windows - II. Primary
  beam model and direction-dependent calibration}.
\newblock {\em \mnras}, 462:4482--4494, November 2016.

\bibitem{asad15}
K.~M.~B. {Asad}, L.~V.~E. {Koopmans}, V.~{Jeli{\'c}}, V.~N. {Pandey},
  A.~{Ghosh}, F.~B. {Abdalla}, G.~{Bernardi}, M.~A. {Brentjens}, A.~G. {de
  Bruyn}, S.~{Bus}, B.~{Ciardi}, E.~{Chapman}, S.~{Daiboo}, E.~R. {Fernandez},
  G.~{Harker}, I.~T. {Iliev}, H.~{Jensen}, O.~{Martinez-Rubi}, G.~{Mellema},
  M.~{Mevius}, A.~R. {Offringa}, A.~H. {Patil}, J.~{Schaye}, R.~M. {Thomas},
  S.~{van der Tol}, H.~K. {Vedantham}, S.~{Yatawatta}, and S.~{Zaroubi}.
\newblock {Polarization leakage in epoch of reionization windows - I. Low
  Frequency Array observations of the 3C196 field}.
\newblock {\em \mnras}, 451:3709--3727, August 2015.

\bibitem{barry16}
N.~{Barry}, B.~{Hazelton}, I.~{Sullivan}, M.~F. {Morales}, and J.~C. {Pober}.
\newblock {Calibration requirements for detecting the 21 cm epoch of
  reionization power spectrum and implications for the SKA}.
\newblock {\em \mnras}, 461:3135--3144, September 2016.

\bibitem{beardsley16}
A.~P. {Beardsley}, B.~J. {Hazelton}, I.~S. {Sullivan}, P.~{Carroll},
  N.~{Barry}, M.~{Rahimi}, B.~{Pindor}, C.~M. {Trott}, J.~{Line}, D.~C.
  {Jacobs}, M.~F. {Morales}, J.~C. {Pober}, G.~{Bernardi}, J.~D. {Bowman},
  M.~P. {Busch}, F.~{Briggs}, R.~J. {Cappallo}, B.~E. {Corey}, A.~{de
  Oliveira-Costa}, J.~S. {Dillon}, D.~{Emrich}, A.~{Ewall-Wice}, L.~{Feng},
  B.~M. {Gaensler}, R.~{Goeke}, L.~J. {Greenhill}, J.~N. {Hewitt},
  N.~{Hurley-Walker}, M.~{Johnston-Hollitt}, D.~L. {Kaplan}, J.~C. {Kasper},
  H.~S. {Kim}, E.~{Kratzenberg}, E.~{Lenc}, A.~{Loeb}, C.~J. {Lonsdale}, M.~J.
  {Lynch}, B.~{McKinley}, S.~R. {McWhirter}, D.~A. {Mitchell}, E.~{Morgan},
  A.~R. {Neben}, N.~{Thyagarajan}, D.~{Oberoi}, A.~R. {Offringa}, S.~M. {Ord},
  S.~{Paul}, T.~{Prabu}, P.~{Procopio}, J.~{Riding}, A.~E.~E. {Rogers},
  A.~{Roshi}, N.~{Udaya Shankar}, S.~K. {Sethi}, K.~S. {Srivani},
  R.~{Subrahmanyan}, M.~{Tegmark}, S.~J. {Tingay}, M.~{Waterson}, R.~B.
  {Wayth}, R.~L. {Webster}, A.~R. {Whitney}, A.~{Williams}, C.~L. {Williams},
  C.~{Wu}, and J.~S.~B. {Wyithe}.
\newblock {First Season MWA EoR Power spectrum Results at Redshift 7}.
\newblock {\em \apj}, 833:102, December 2016.

\bibitem{bernardi09}
G.~{Bernardi}, A.~G. {de Bruyn}, M.~A. {Brentjens}, B.~{Ciardi}, G.~{Harker},
  V.~{Jeli{\'c}}, L.~V.~E. {Koopmans}, P.~{Labropoulos}, A.~{Offringa}, V.~N.
  {Pandey}, J.~{Schaye}, R.~M. {Thomas}, S.~{Yatawatta}, and S.~{Zaroubi}.
\newblock {Foregrounds for observations of the cosmological 21 cm line. I.
  First Westerbork measurements of Galactic emission at 150 MHz in a low
  latitude field}.
\newblock {\em \aap}, 500:965--979, June 2009.

\bibitem{bernardi10}
G.~{Bernardi}, A.~G. {de Bruyn}, G.~{Harker}, M.~A. {Brentjens}, B.~{Ciardi},
  V.~{Jeli{\'c}}, L.~V.~E. {Koopmans}, P.~{Labropoulos}, A.~{Offringa}, V.~N.
  {Pandey}, J.~{Schaye}, R.~M. {Thomas}, S.~{Yatawatta}, and S.~{Zaroubi}.
\newblock {Foregrounds for observations of the cosmological 21 cm line. II.
  Westerbork observations of the fields around 3C 196 and the North Celestial
  Pole}.
\newblock {\em \aap}, 522:A67, November 2010.

\bibitem{bernardi13}
G.~{Bernardi}, L.~J. {Greenhill}, D.~A. {Mitchell}, S.~M. {Ord}, B.~J.
  {Hazelton}, B.~M. {Gaensler}, A.~{de Oliveira-Costa}, M.~F. {Morales},
  N.~{Udaya Shankar}, R.~{Subrahmanyan}, R.~B. {Wayth}, E.~{Lenc}, C.~L.
  {Williams}, W.~{Arcus}, B.~S. {Arora}, D.~G. {Barnes}, J.~D. {Bowman}, F.~H.
  {Briggs}, J.~D. {Bunton}, R.~J. {Cappallo}, B.~E. {Corey}, A.~{Deshpande},
  L.~{deSouza}, D.~{Emrich}, R.~{Goeke}, D.~{Herne}, J.~N. {Hewitt},
  M.~{Johnston-Hollitt}, D.~{Kaplan}, J.~C. {Kasper}, B.~B. {Kincaid},
  R.~{Koenig}, E.~{Kratzenberg}, C.~J. {Lonsdale}, M.~J. {Lynch}, S.~R.
  {McWhirter}, E.~{Morgan}, D.~{Oberoi}, J.~{Pathikulangara}, T.~{Prabu}, R.~A.
  {Remillard}, A.~E.~E. {Rogers}, A.~{Roshi}, J.~E. {Salah}, R.~J. {Sault},
  K.~S. {Srivani}, J.~{Stevens}, S.~J. {Tingay}, M.~{Waterson}, R.~L.
  {Webster}, A.~R. {Whitney}, A.~{Williams}, and J.~S.~B. {Wyithe}.
\newblock {A 189 MHz, 2400 deg$^{2}$ Polarization Survey with the Murchison
  Widefield Array 32-element Prototype}.
\newblock {\em \apj}, 771:105, July 2013.

\bibitem{bernardi11}
G.~{Bernardi}, D.~A. {Mitchell}, S.~M. {Ord}, L.~J. {Greenhill}, B.~{Pindor},
  R.~B. {Wayth}, and J.~S.~B. {Wyithe}.
\newblock {Subtraction of point sources from interferometric radio images
  through an algebraic forward modelling scheme}.
\newblock {\em \mnras}, 413:411--422, May 2011.

\bibitem{bhatnagar08}
S.~{Bhatnagar}, T.~J. {Cornwell}, K.~{Golap}, and J.~M. {Uson}.
\newblock {Correcting direction-dependent gains in the deconvolution of radio
  interferometric images}.
\newblock {\em \aap}, 487:419--429, August 2008.

\bibitem{byrne19}
R.~{Byrne}, M.~F. {Morales}, B.~{Hazelton}, W.~{Li}, N.~{Barry}, A.~P.
  {Beardsley}, R.~{Joseph}, J.~{Pober}, I.~{Sullivan}, and C.~{Trott}.
\newblock {Fundamental Limitations on the Calibration of Redundant 21 cm
  Cosmology Instruments and Implications for HERA and the SKA}.
\newblock {\em \apj}, 875:70, April 2019.

\bibitem{carroll16}
P.~A. {Carroll}, J.~{Line}, M.~F. {Morales}, N.~{Barry}, A.~P. {Beardsley},
  B.~J. {Hazelton}, D.~C. {Jacobs}, J.~C. {Pober}, I.~S. {Sullivan}, R.~L.
  {Webster}, G.~{Bernardi}, J.~D. {Bowman}, F.~{Briggs}, R.~J. {Cappallo},
  B.~E. {Corey}, A.~{de Oliveira-Costa}, J.~S. {Dillon}, D.~{Emrich},
  A.~{Ewall-Wice}, L.~{Feng}, B.~M. {Gaensler}, R.~{Goeke}, L.~J. {Greenhill},
  J.~N. {Hewitt}, N.~{Hurley-Walker}, M.~{Johnston-Hollitt}, D.~L. {Kaplan},
  J.~C. {Kasper}, H.~{Kim}, E.~{Kratzenberg}, E.~{Lenc}, A.~{Loeb}, C.~J.
  {Lonsdale}, M.~J. {Lynch}, B.~{McKinley}, S.~R. {McWhirter}, D.~A.
  {Mitchell}, E.~{Morgan}, A.~R. {Neben}, D.~{Oberoi}, A.~R. {Offringa}, S.~M.
  {Ord}, S.~{Paul}, B.~{Pindor}, T.~{Prabu}, P.~{Procopio}, J.~{Riding},
  A.~E.~E. {Rogers}, A.~{Roshi}, N.~U. {Shankar}, S.~K. {Sethi}, K.~S.
  {Srivani}, R.~{Subrahmanyan}, M.~{Tegmark}, N.~{Thyagarajan}, S.~J. {Tingay},
  C.~M. {Trott}, M.~{Waterson}, R.~B. {Wayth}, A.~R. {Whitney}, A.~{Williams},
  C.~L. {Williams}, C.~{Wu}, and J.~S.~B. {Wyithe}.
\newblock {A high reliability survey of discrete Epoch of Reionization
  foreground sources in the MWA EoR0 field}.
\newblock {\em \mnras}, 461:4151--4175, October 2016.

\bibitem{choudhuri17}
S.~{Choudhuri}, S.~{Bharadwaj}, S.~S. {Ali}, N.~{Roy}, H.~T. {Intema}, and
  A.~{Ghosh}.
\newblock {The angular power spectrum measurement of the Galactic synchrotron
  emission in two fields of the TGSS survey}.
\newblock {\em \mnras}, 470:L11--L15, September 2017.

\bibitem{cornwell81}
T.~J. {Cornwell} and P.~N. {Wilkinson}.
\newblock {A new method for making maps with unstable radio interferometers}.
\newblock {\em \mnras}, 196:1067--1086, September 1981.

\bibitem{datta12}
K.~K. {Datta}, M.~M. {Friedrich}, G.~{Mellema}, I.~T. {Iliev}, and P.~R.
  {Shapiro}.
\newblock {Prospects of observing a quasar H II region during the epoch of
  reionization with the redshifted 21-cm signal}.
\newblock {\em \mnras}, 424:762--778, July 2012.

\bibitem{deleraacedo18}
E.~{de Lera Acedo}, P.~{Bolli}, F.~{Paonessa}, G.~{Virone}, E.~{Colin-Beltran},
  N.~{Razavi-Ghods}, I.~{Aicardi}, A.~{Lingua}, P.~{Maschio}, J.~{Monari},
  G.~{Naldi}, M.~{Piras}, and G.~{Pupillo}.
\newblock {SKA aperture array verification system: electromagnetic modeling and
  beam pattern measurements using a micro UAV}.
\newblock {\em Experimental Astronomy}, 45:1--20, March 2018.

\bibitem{deleraacedo17}
E.~{de Lera Acedo}, C.~M. {Trott}, R.~B. {Wayth}, N.~{Fagnoni}, G.~{Bernardi},
  B.~{Wakley}, L.~V.~E. {Koopmans}, A.~J. {Faulkner}, and J.~G. {bij de Vaate}.
\newblock {Spectral performance of SKA Log-periodic Antennas I: mitigating
  spectral artefacts in SKA1-LOW 21 cm cosmology experiments}.
\newblock {\em \mnras}, 469:2662--2671, August 2017.

\bibitem{dickinson04}
C.~{Dickinson}, R.~A. {Battye}, P.~{Carreira}, K.~{Cleary}, R.~D. {Davies},
  R.~J. {Davis}, R.~{Genova-Santos}, K.~{Grainge}, C.~M. {Guti{\'e}rrez}, Y.~A.
  {Hafez}, M.~P. {Hobson}, M.~E. {Jones}, R.~{Kneissl}, K.~{Lancaster},
  A.~{Lasenby}, J.~P. {Leahy}, K.~{Maisinger}, C.~{{\"O}dman}, G.~{Pooley},
  N.~{Rajguru}, R.~{Rebolo}, J.~A. {Rubi{\~n}o-Martin}, R.~D.~E. {Saunders},
  R.~S. {Savage}, A.~{Scaife}, P.~F. {Scott}, A.~{Slosar}, P.~{Sosa Molina},
  A.~C. {Taylor}, D.~{Titterington}, E.~{Waldram}, R.~A. {Watson}, and
  A.~{Wilkinson}.
\newblock {High-sensitivity measurements of the cosmic microwave background
  power spectrum with the extended Very Small Array}.
\newblock {\em \mnras}, 353:732--746, September 2004.

\bibitem{dillon18}
J.~S. {Dillon}, S.~A. {Kohn}, A.~R. {Parsons}, J.~E. {Aguirre}, Z.~S. {Ali},
  G.~{Bernardi}, N.~S. {Kern}, W.~{Li}, A.~{Liu}, C.~D. {Nunhokee}, and J.~C.
  {Pober}.
\newblock {Polarized redundant-baseline calibration for 21 cm cosmology without
  adding spectral structure}.
\newblock {\em \mnras}, 477:5670--5681, July 2018.

\bibitem{dowell17}
J.~{Dowell}, G.~B. {Taylor}, F.~K. {Schinzel}, N.~E. {Kassim}, and
  K.~{Stovall}.
\newblock {The LWA1 Low Frequency Sky Survey}.
\newblock {\em \mnras}, 469:4537--4550, August 2017.

\bibitem{ewall-wice17}
A.~{Ewall-Wice}, J.~S. {Dillon}, A.~{Liu}, and J.~{Hewitt}.
\newblock {The impact of modelling errors on interferometer calibration for 21
  cm power spectra}.
\newblock {\em \mnras}, 470:1849--1870, September 2017.

\bibitem{geil11}
P.~M. {Geil}, B.~M. {Gaensler}, and J.~S.~B. {Wyithe}.
\newblock {Polarized foreground removal at low radio frequencies using rotation
  measure synthesis: uncovering the signature of hydrogen reionization}.
\newblock {\em \mnras}, 418:516--535, November 2011.

\bibitem{grobler14}
T.~L. {Grobler}, C.~D. {Nunhokee}, O.~M. {Smirnov}, A.~J. {van Zyl}, and A.~G.
  {de Bruyn}.
\newblock {Calibration artefacts in radio interferometry - I. Ghost sources in
  Westerbork Synthesis Radio Telescope data}.
\newblock {\em \mnras}, 439:4030--4047, April 2014.

\bibitem{grobler16}
T.~L. {Grobler}, A.~J. {Stewart}, S.~J. {Wijnholds}, J.~S. {Kenyon}, and O.~M.
  {Smirnov}.
\newblock {Calibration artefacts in radio interferometry - III. Phase-only
  calibration and primary beam correction}.
\newblock {\em \mnras}, 461:2975--2992, September 2016.

\bibitem{halverson02}
N.~W. {Halverson}, E.~M. {Leitch}, C.~{Pryke}, J.~{Kovac}, J.~E. {Carlstrom},
  W.~L. {Holzapfel}, M.~{Dragovan}, J.~K. {Cartwright}, B.~S. {Mason},
  S.~{Padin}, T.~J. {Pearson}, A.~C.~S. {Readhead}, and M.~C. {Shepherd}.
\newblock {Degree Angular Scale Interferometer First Results: A Measurement of
  the Cosmic Microwave Background Angular Power Spectrum}.
\newblock {\em \apj}, 568:38--45, March 2002.

\bibitem{hamaker96}
J.~P. {Hamaker}, J.~D. {Bregman}, and R.~J. {Sault}.
\newblock {Understanding radio polarimetry. I. Mathematical foundations.}
\newblock {\em \aaps}, 117:137--147, May 1996.

\bibitem{harker10}
G.~{Harker}, S.~{Zaroubi}, G.~{Bernardi}, M.~A. {Brentjens}, A.~G. {de Bruyn},
  B.~{Ciardi}, V.~{Jeli{\'c}}, L.~V.~E. {Koopmans}, P.~{Labropoulos},
  G.~{Mellema}, A.~{Offringa}, V.~N. {Pandey}, A.~H. {Pawlik}, J.~{Schaye},
  R.~M. {Thomas}, and S.~{Yatawatta}.
\newblock {Power spectrum extraction for redshifted 21-cm Epoch of Reionization
  experiments: the LOFAR case}.
\newblock {\em \mnras}, 405:2492--2504, July 2010.

\bibitem{hurley-walker17}
N.~{Hurley-Walker}, J.~R. {Callingham}, P.~J. {Hancock}, T.~M.~O. {Franzen},
  L.~{Hindson}, A.~D. {Kapi{\'n}ska}, J.~{Morgan}, A.~R. {Offringa}, R.~B.
  {Wayth}, C.~{Wu}, Q.~{Zheng}, T.~{Murphy}, M.~E. {Bell}, K.~S. {Dwarakanath},
  B.~{For}, B.~M. {Gaensler}, M.~{Johnston-Hollitt}, E.~{Lenc}, P.~{Procopio},
  L.~{Staveley-Smith}, R.~{Ekers}, J.~D. {Bowman}, F.~{Briggs}, R.~J.
  {Cappallo}, A.~A. {Deshpande}, L.~{Greenhill}, B.~J. {Hazelton}, D.~L.
  {Kaplan}, C.~J. {Lonsdale}, S.~R. {McWhirter}, D.~A. {Mitchell}, M.~F.
  {Morales}, E.~{Morgan}, D.~{Oberoi}, S.~M. {Ord}, T.~{Prabu}, N.~U.
  {Shankar}, K.~S. {Srivani}, R.~{Subrahmanyan}, S.~J. {Tingay}, R.~L.
  {Webster}, A.~{Williams}, and C.~L. {Williams}.
\newblock {GaLactic and Extragalactic All-sky Murchison Widefield Array (GLEAM)
  survey - I. A low-frequency extragalactic catalogue}.
\newblock {\em \mnras}, 464:1146--1167, January 2017.

\bibitem{iacobelli13}
M.~{Iacobelli}, M.~{Haverkorn}, E.~{Orr{\'u}}, R.~F. {Pizzo}, J.~{Anderson},
  R.~{Beck}, M.~R. {Bell}, A.~{Bonafede}, K.~{Chyzy}, R.-J. {Dettmar}, T.~A.
  {En{\ss}lin}, G.~{Heald}, C.~{Horellou}, A.~{Horneffer}, W.~{Jurusik},
  H.~{Junklewitz}, M.~{Kuniyoshi}, D.~D. {Mulcahy}, R.~{Paladino}, W.~{Reich},
  A.~{Scaife}, C.~{Sobey}, C.~{Sotomayor-Beltran}, A.~{Alexov}, A.~{Asgekar},
  I.~M. {Avruch}, M.~E. {Bell}, I.~{van Bemmel}, M.~J. {Bentum}, G.~{Bernardi},
  P.~{Best}, L.~{B{\i}rzan}, F.~{Breitling}, J.~{Broderick}, W.~N. {Brouw},
  M.~{Br{\"u}ggen}, H.~R. {Butcher}, B.~{Ciardi}, J.~E. {Conway}, F.~{de
  Gasperin}, E.~{de Geus}, S.~{Duscha}, J.~{Eisl{\"o}ffel}, D.~{Engels},
  H.~{Falcke}, R.~A. {Fallows}, C.~{Ferrari}, W.~{Frieswijk}, M.~A. {Garrett},
  J.~{Grie{\ss}meier}, A.~W. {Gunst}, J.~P. {Hamaker}, T.~E. {Hassall},
  J.~W.~T. {Hessels}, M.~{Hoeft}, J.~{H{\"o}randel}, V.~{Jelic},
  A.~{Karastergiou}, V.~I. {Kondratiev}, L.~V.~E. {Koopmans}, M.~{Kramer},
  G.~{Kuper}, J.~{van Leeuwen}, G.~{Macario}, G.~{Mann}, J.~P. {McKean},
  H.~{Munk}, M.~{Pandey-Pommier}, A.~G. {Polatidis}, H.~{R{\"o}ttgering},
  D.~{Schwarz}, J.~{Sluman}, O.~{Smirnov}, B.~W. {Stappers}, M.~{Steinmetz},
  M.~{Tagger}, Y.~{Tang}, C.~{Tasse}, C.~{Toribio}, R.~{Vermeulen}, C.~{Vocks},
  C.~{Vogt}, R.~J. {van Weeren}, M.~W. {Wise}, O.~{Wucknitz}, S.~{Yatawatta},
  P.~{Zarka}, and A.~{Zensus}.
\newblock {Studying Galactic interstellar turbulence through fluctuations in
  synchrotron emission. First LOFAR Galactic foreground detection}.
\newblock {\em \aap}, 558:A72, October 2013.

\bibitem{intema17}
H.~T. {Intema}, P.~{Jagannathan}, K.~P. {Mooley}, and D.~A. {Frail}.
\newblock {The GMRT 150 MHz all-sky radio survey. First alternative data
  release TGSS ADR1}.
\newblock {\em \aap}, 598:A78, February 2017.

\bibitem{intema09}
H.~T. {Intema}, S.~{van der Tol}, W.~D. {Cotton}, A.~S. {Cohen}, I.~M. {van
  Bemmel}, and H.~J.~A. {R{\"o}ttgering}.
\newblock {Ionospheric calibration of low frequency radio interferometric
  observations using the peeling scheme. I. Method description and first
  results}.
\newblock {\em \aap}, 501:1185--1205, July 2009.

\bibitem{jacobs11}
D.~C. {Jacobs}, J.~E. {Aguirre}, A.~R. {Parsons}, J.~C. {Pober}, R.~F.
  {Bradley}, C.~L. {Carilli}, N.~E. {Gugliucci}, J.~R. {Manley}, C.~{van der
  Merwe}, D.~F. {Moore}, and C.~R. {Parashare}.
\newblock {New 145 MHz Source Measurements by PAPER in the Southern Sky}.
\newblock {\em \apjl}, 734:L34, June 2011.

\bibitem{jacobs17}
D.~C. {Jacobs}, J.~{Burba}, J.~D. {Bowman}, A.~R. {Neben}, B.~{Stinnett},
  L.~{Turner}, K.~{Johnson}, M.~{Busch}, J.~{Allison}, M.~{Leatham},
  V.~{Serrano Rodriguez}, M.~{Denney}, and D.~{Nelson}.
\newblock {First Demonstration of ECHO: an External Calibrator for Hydrogen
  Observatories}.
\newblock {\em \pasp}, 129(3):035002, March 2017.

\bibitem{jacobs15}
D.~C. {Jacobs}, J.~C. {Pober}, A.~R. {Parsons}, J.~E. {Aguirre}, Z.~S. {Ali},
  J.~{Bowman}, R.~F. {Bradley}, C.~L. {Carilli}, D.~R. {DeBoer}, M.~R.
  {Dexter}, N.~E. {Gugliucci}, P.~{Klima}, A.~{Liu}, D.~H.~E. {MacMahon}, J.~R.
  {Manley}, D.~F. {Moore}, I.~I. {Stefan}, and W.~P. {Walbrugh}.
\newblock {Multiredshift Limits on the 21 cm Power Spectrum from PAPER}.
\newblock {\em \apj}, 801:51, March 2015.

\bibitem{jelic15}
V.~{Jeli{\'c}}, A.~G. {de Bruyn}, V.~N. {Pandey}, M.~{Mevius}, M.~{Haverkorn},
  M.~A. {Brentjens}, L.~V.~E. {Koopmans}, S.~{Zaroubi}, F.~B. {Abdalla},
  K.~M.~B. {Asad}, S.~{Bus}, E.~{Chapman}, B.~{Ciardi}, E.~R. {Fernandez},
  A.~{Ghosh}, G.~{Harker}, I.~T. {Iliev}, H.~{Jensen}, S.~{Kazemi},
  G.~{Mellema}, A.~R. {Offringa}, A.~H. {Patil}, H.~K. {Vedantham}, and
  S.~{Yatawatta}.
\newblock {Linear polarization structures in LOFAR observations of the
  interstellar medium in the 3C 196 field}.
\newblock {\em \aap}, 583:A137, November 2015.

\bibitem{jelic10}
V.~{Jeli{\'c}}, S.~{Zaroubi}, P.~{Labropoulos}, G.~{Bernardi}, A.~G. {de
  Bruyn}, and L.~V.~E. {Koopmans}.
\newblock {Realistic simulations of the Galactic polarized foreground:
  consequences for 21-cm reionization detection experiments}.
\newblock {\em \mnras}, 409:1647--1659, December 2010.

\bibitem{jordan17}
C.~H. {Jordan}, S.~{Murray}, C.~M. {Trott}, R.~B. {Wayth}, D.~A. {Mitchell},
  M.~{Rahimi}, B.~{Pindor}, P.~{Procopio}, and J.~{Morgan}.
\newblock {Characterization of the ionosphere above the Murchison Radio
  Observatory using the Murchison Widefield Array}.
\newblock {\em \mnras}, 471:3974--3987, November 2017.

\bibitem{joseph18}
R.~C. {Joseph}, C.~M. {Trott}, and R.~B. {Wayth}.
\newblock {The Bias and Uncertainty of Redundant and Sky-based Calibration
  Under Realistic Sky and Telescope Conditions}.
\newblock {\em \aj}, 156:285, December 2018.

\bibitem{kakiichi17}
K.~{Kakiichi}, S.~{Majumdar}, G.~{Mellema}, B.~{Ciardi}, K.~L. {Dixon}, I.~T.
  {Iliev}, V.~{Jeli{\'c}}, L.~V.~E. {Koopmans}, S.~{Zaroubi}, and P.~{Busch}.
\newblock {Recovering the H II region size statistics from 21-cm tomography}.
\newblock {\em \mnras}, 471:1936--1954, October 2017.

\bibitem{kazemi11}
S.~{Kazemi}, S.~{Yatawatta}, S.~{Zaroubi}, P.~{Lampropoulos}, A.~G. {de Bruyn},
  L.~V.~E. {Koopmans}, and J.~{Noordam}.
\newblock {Radio interferometric calibration using the SAGE algorithm}.
\newblock {\em \mnras}, 414:1656--1666, June 2011.

\bibitem{lenc16}
E.~{Lenc}, B.~M. {Gaensler}, X.~H. {Sun}, E.~M. {Sadler}, A.~G. {Willis},
  N.~{Barry}, A.~P. {Beardsley}, M.~E. {Bell}, G.~{Bernardi}, J.~D. {Bowman},
  F.~{Briggs}, J.~R. {Callingham}, R.~J. {Cappallo}, P.~{Carroll}, B.~E.
  {Corey}, A.~{de Oliveira-Costa}, A.~A. {Deshpande}, J.~S. {Dillon}, K.~S.
  {Dwarkanath}, D.~{Emrich}, A.~{Ewall-Wice}, L.~{Feng}, B.-Q. {For},
  R.~{Goeke}, L.~J. {Greenhill}, P.~{Hancock}, B.~J. {Hazelton}, J.~N.
  {Hewitt}, L.~{Hindson}, N.~{Hurley-Walker}, M.~{Johnston-Hollitt}, D.~C.
  {Jacobs}, A.~D. {Kapi{\'n}ska}, D.~L. {Kaplan}, J.~C. {Kasper}, H.-S. {Kim},
  E.~{Kratzenberg}, J.~{Line}, A.~{Loeb}, C.~J. {Lonsdale}, M.~J. {Lynch},
  B.~{McKinley}, S.~R. {McWhirter}, D.~A. {Mitchell}, M.~F. {Morales},
  E.~{Morgan}, J.~{Morgan}, T.~{Murphy}, A.~R. {Neben}, D.~{Oberoi}, A.~R.
  {Offringa}, S.~M. {Ord}, S.~{Paul}, B.~{Pindor}, J.~C. {Pober}, T.~{Prabu},
  P.~{Procopio}, J.~{Riding}, A.~E.~E. {Rogers}, A.~{Roshi}, N.~{Udaya
  Shankar}, S.~K. {Sethi}, K.~S. {Srivani}, L.~{Staveley-Smith},
  R.~{Subrahmanyan}, I.~S. {Sullivan}, M.~{Tegmark}, N.~{Thyagarajan}, S.~J.
  {Tingay}, C.~{Trott}, M.~{Waterson}, R.~B. {Wayth}, R.~L. {Webster}, A.~R.
  {Whitney}, A.~{Williams}, C.~L. {Williams}, C.~{Wu}, J.~S.~B. {Wyithe}, and
  Q.~{Zheng}.
\newblock {Low-frequency Observations of Linearly Polarized Structures in the
  Interstellar Medium near the South Galactic Pole}.
\newblock {\em \apj}, 830:38, October 2016.

\bibitem{li18}
W.~{Li}, J.~C. {Pober}, B.~J. {Hazelton}, N.~{Barry}, M.~F. {Morales},
  I.~{Sullivan}, A.~R. {Parsons}, Z.~S. {Ali}, J.~S. {Dillon}, A.~P.
  {Beardsley}, J.~D. {Bowman}, F.~{Briggs}, R.~{Byrne}, P.~{Carroll},
  B.~{Crosse}, D.~{Emrich}, A.~{Ewall-Wice}, L.~{Feng}, T.~M.~O. {Franzen},
  J.~N. {Hewitt}, L.~{Horsley}, D.~C. {Jacobs}, M.~{Johnston-Hollitt},
  C.~{Jordan}, R.~C. {Joseph}, D.~L. {Kaplan}, D.~{Kenney}, H.~{Kim},
  P.~{Kittiwisit}, A.~{Lanman}, J.~{Line}, B.~{McKinley}, D.~A. {Mitchell},
  S.~{Murray}, A.~{Neben}, A.~R. {Offringa}, D.~{Pallot}, S.~{Paul},
  B.~{Pindor}, P.~{Procopio}, M.~{Rahimi}, J.~{Riding}, S.~K. {Sethi},
  N.~{Udaya Shankar}, K.~{Steele}, R.~{Subrahmanian}, M.~{Tegmark},
  N.~{Thyagarajan}, S.~J. {Tingay}, C.~{Trott}, M.~{Walker}, R.~B. {Wayth},
  R.~L. {Webster}, A.~{Williams}, C.~{Wu}, and S.~{Wyithe}.
\newblock {Comparing Redundant and Sky-model-based Interferometric Calibration:
  A First Look with Phase II of the MWA}.
\newblock {\em \apj}, 863:170, August 2018.

\bibitem{liu10}
A.~{Liu}, M.~{Tegmark}, S.~{Morrison}, A.~{Lutomirski}, and M.~{Zaldarriaga}.
\newblock {Precision calibration of radio interferometers using redundant
  baselines}.
\newblock {\em \mnras}, 408:1029--1050, October 2010.

\bibitem{majumdar12}
S.~{Majumdar}, S.~{Bharadwaj}, and T.~R. {Choudhury}.
\newblock {Constrainingquasar and intergalactic medium properties through
  bubble detection in redshifted 21-cm maps}.
\newblock {\em \mnras}, 426:3178--3194, November 2012.

\bibitem{mellema13}
G.~{Mellema}, L.~V.~E. {Koopmans}, F.~A. {Abdalla}, G.~{Bernardi}, B.~{Ciardi},
  S.~{Daiboo}, A.~G. {de Bruyn}, K.~K. {Datta}, H.~{Falcke}, A.~{Ferrara},
  I.~T. {Iliev}, F.~{Iocco}, V.~{Jeli{\'c}}, H.~{Jensen}, R.~{Joseph},
  P.~{Labroupoulos}, A.~{Meiksin}, A.~{Mesinger}, A.~R. {Offringa}, V.~N.
  {Pandey}, J.~R. {Pritchard}, M.~G. {Santos}, D.~J. {Schwarz}, B.~{Semelin},
  H.~{Vedantham}, S.~{Yatawatta}, and S.~{Zaroubi}.
\newblock {Reionization and the Cosmic Dawn with the Square Kilometre Array}.
\newblock {\em Experimental Astronomy}, 36:235--318, August 2013.

\bibitem{mitchell08}
D.~A. {Mitchell}, L.~J. {Greenhill}, R.~B. {Wayth}, R.~J. {Sault}, C.~J.
  {Lonsdale}, R.~J. {Cappallo}, M.~F. {Morales}, and S.~M. {Ord}.
\newblock {Real-Time Calibration of the Murchison Widefield Array}.
\newblock {\em IEEE Journal of Selected Topics in Signal Processing},
  2:707--717, November 2008.

\bibitem{moore13}
D.~F. {Moore}, J.~E. {Aguirre}, A.~R. {Parsons}, D.~C. {Jacobs}, and J.~C.
  {Pober}.
\newblock {The Effects of Polarized Foregrounds on 21 cm Epoch of Reionization
  Power Spectrum Measurements}.
\newblock {\em \apj}, 769:154, June 2013.

\bibitem{morales05}
M.~F. {Morales}.
\newblock {Power Spectrum Sensitivity and the Design of Epoch of Reionization
  Observatories}.
\newblock {\em \apj}, 619:678--683, February 2005.

\bibitem{morales12}
M.~F. {Morales}, B.~{Hazelton}, I.~{Sullivan}, and A.~{Beardsley}.
\newblock {Four Fundamental Foreground Power Spectrum Shapes for 21 cm
  Cosmology Observations}.
\newblock {\em \apj}, 752:137, June 2012.

\bibitem{morales04}
M.~F. {Morales} and J.~{Hewitt}.
\newblock {Toward Epoch of Reionization Measurements with Wide-Field Radio
  Observations}.
\newblock {\em \apj}, 615:7--18, November 2004.

\bibitem{sardabaradi19}
A.~{Mouri Sardarabadi} and L.~V.~E. {Koopmans}.
\newblock {Quantifying suppression of the cosmological 21-cm signal due to
  direction-dependent gain calibration in radio interferometers}.
\newblock {\em \mnras}, 483:5480--5490, March 2019.

\bibitem{neben16}
A.~R. {Neben}, J.~N. {Hewitt}, R.~F. {Bradley}, J.~S. {Dillon}, G.~{Bernardi},
  J.~D. {Bowman}, F.~{Briggs}, R.~J. {Cappallo}, B.~E. {Corey}, A.~A.
  {Deshpande}, R.~{Goeke}, L.~J. {Greenhill}, B.~J. {Hazelton},
  M.~{Johnston-Hollitt}, D.~L. {Kaplan}, C.~J. {Lonsdale}, S.~R. {McWhirter},
  D.~A. {Mitchell}, M.~F. {Morales}, E.~{Morgan}, D.~{Oberoi}, S.~M. {Ord},
  T.~{Prabu}, N.~{Udaya Shankar}, K.~S. {Srivani}, R.~{Subrahmanyan}, S.~J.
  {Tingay}, R.~B. {Wayth}, R.~L. {Webster}, A.~{Williams}, and C.~L.
  {Williams}.
\newblock {Beam-forming Errors in Murchison Widefield Array Phased Array
  Antennas and their Effects on Epoch of Reionization Science}.
\newblock {\em \apj}, 820:44, March 2016.

\bibitem{nunhokee17}
C.~D. {Nunhokee}, G.~{Bernardi}, S.~A. {Kohn}, J.~E. {Aguirre},
  N.~{Thyagarajan}, J.~S. {Dillon}, G.~{Foster}, T.~L. {Grobler}, J.~Z.~E.
  {Martinot}, and A.~R. {Parsons}.
\newblock {Constraining Polarized Foregrounds for EoR Experiments. II.
  Polarization Leakage Simulations in the Avoidance Scheme}.
\newblock {\em \apj}, 848:47, October 2017.

\bibitem{orosz19}
N.~{Orosz}, J.~S. {Dillon}, A.~{Ewall-Wice}, A.~R. {Parsons}, and
  N.~{Thyagarajan}.
\newblock {Mitigating the effects of antenna-to-antenna variation on
  redundant-baseline calibration for 21 cm cosmology}.
\newblock {\em \mnras}, 487:537--549, July 2019.

\bibitem{parsons12b}
A.~{Parsons}, J.~{Pober}, M.~{McQuinn}, D.~{Jacobs}, and J.~{Aguirre}.
\newblock {A Sensitivity and Array-configuration Study for Measuring the Power
  Spectrum of 21 cm Emission from Reionization}.
\newblock {\em \apj}, 753:81, July 2012.

\bibitem{parsons10}
A.~R. {Parsons}, D.~C. {Backer}, G.~S. {Foster}, M.~C.~H. {Wright}, R.~F.
  {Bradley}, N.~E. {Gugliucci}, C.~R. {Parashare}, E.~E. {Benoit}, J.~E.
  {Aguirre}, D.~C. {Jacobs}, C.~L. {Carilli}, D.~{Herne}, M.~J. {Lynch}, J.~R.
  {Manley}, and D.~J. {Werthimer}.
\newblock {The Precision Array for Probing the Epoch of Re-ionization: Eight
  Station Results}.
\newblock {\em \aj}, 139:1468--1480, April 2010.

\bibitem{parsons14}
A.~R. {Parsons}, A.~{Liu}, J.~E. {Aguirre}, Z.~S. {Ali}, R.~F. {Bradley}, C.~L.
  {Carilli}, D.~R. {DeBoer}, M.~R. {Dexter}, N.~E. {Gugliucci}, D.~C. {Jacobs},
  P.~{Klima}, D.~H.~E. {MacMahon}, J.~R. {Manley}, D.~F. {Moore}, J.~C.
  {Pober}, I.~I. {Stefan}, and W.~P. {Walbrugh}.
\newblock {New Limits on 21 cm Epoch of Reionization from PAPER-32 Consistent
  with an X-Ray Heated Intergalactic Medium at z = 7.7}.
\newblock {\em \apj}, 788:106, June 2014.

\bibitem{parsons12a}
A.~R. {Parsons}, J.~C. {Pober}, J.~E. {Aguirre}, C.~L. {Carilli}, D.~C.
  {Jacobs}, and D.~F. {Moore}.
\newblock {A Per-baseline, Delay-spectrum Technique for Accessing the 21 cm
  Cosmic Reionization Signature}.
\newblock {\em \apj}, 756:165, September 2012.

\bibitem{patil17}
A.~H. {Patil}, S.~{Yatawatta}, L.~V.~E. {Koopmans}, A.~G. {de Bruyn}, M.~A.
  {Brentjens}, S.~{Zaroubi}, K.~M.~B. {Asad}, M.~{Hatef}, V.~{Jeli{\'c}},
  M.~{Mevius}, A.~R. {Offringa}, V.~N. {Pandey}, H.~{Vedantham}, F.~B.
  {Abdalla}, W.~N. {Brouw}, E.~{Chapman}, B.~{Ciardi}, B.~K. {Gehlot},
  A.~{Ghosh}, G.~{Harker}, I.~T. {Iliev}, K.~{Kakiichi}, S.~{Majumdar},
  G.~{Mellema}, M.~B. {Silva}, J.~{Schaye}, D.~{Vrbanec}, and S.~J.
  {Wijnholds}.
\newblock {Upper Limits on the 21 cm Epoch of Reionization Power Spectrum from
  One Night with LOFAR}.
\newblock {\em \apj}, 838:65, March 2017.

\bibitem{patil16}
A.~H. {Patil}, S.~{Yatawatta}, S.~{Zaroubi}, L.~V.~E. {Koopmans}, A.~G. {de
  Bruyn}, V.~{Jeli{\'c}}, B.~{Ciardi}, I.~T. {Iliev}, M.~{Mevius}, V.~N.
  {Pandey}, and B.~K. {Gehlot}.
\newblock {Systematic biases in low-frequency radio interferometric data due to
  calibration: the LOFAR-EoR case}.
\newblock {\em \mnras}, 463:4317--4330, December 2016.

\bibitem{pearson84}
T.~J. {Pearson} and A.~C.~S. {Readhead}.
\newblock {Image Formation by Self-Calibration in Radio Astronomy}.
\newblock {\em \araa}, 22:97--130, 1984.

\bibitem{pen09}
U.-L. {Pen}, T.-C. {Chang}, C.~M. {Hirata}, J.~B. {Peterson}, J.~{Roy},
  Y.~{Gupta}, J.~{Odegova}, and K.~{Sigurdson}.
\newblock {The GMRT EoR experiment: limits on polarized sky brightness at 150
  MHz}.
\newblock {\em \mnras}, 399:181--194, October 2009.

\bibitem{pober16}
J.~C. {Pober}, B.~J. {Hazelton}, A.~P. {Beardsley}, N.~A. {Barry}, Z.~E.
  {Martinot}, I.~S. {Sullivan}, M.~F. {Morales}, M.~E. {Bell}, G.~{Bernardi},
  N.~D.~R. {Bhat}, J.~D. {Bowman}, F.~{Briggs}, R.~J. {Cappallo}, P.~{Carroll},
  B.~E. {Corey}, A.~{de Oliveira-Costa}, A.~A. {Deshpande}, J.~S. {Dillon},
  D.~{Emrich}, A.~M. {Ewall-Wice}, L.~{Feng}, R.~{Goeke}, L.~J. {Greenhill},
  J.~N. {Hewitt}, L.~{Hindson}, N.~{Hurley-Walker}, D.~C. {Jacobs},
  M.~{Johnston-Hollitt}, D.~L. {Kaplan}, J.~C. {Kasper}, H.-S. {Kim},
  P.~{Kittiwisit}, E.~{Kratzenberg}, N.~{Kudryavtseva}, E.~{Lenc}, J.~{Line},
  A.~{Loeb}, C.~J. {Lonsdale}, M.~J. {Lynch}, B.~{McKinley}, S.~R. {McWhirter},
  D.~A. {Mitchell}, E.~{Morgan}, A.~R. {Neben}, D.~{Oberoi}, A.~R. {Offringa},
  S.~M. {Ord}, S.~{Paul}, B.~{Pindor}, T.~{Prabu}, P.~{Procopio}, J.~{Riding},
  A.~E.~E. {Rogers}, A.~{Roshi}, S.~K. {Sethi}, N.~{Udaya Shankar}, K.~S.
  {Srivani}, R.~{Subrahmanyan}, M.~{Tegmark}, N.~{Thyagarajan}, S.~J. {Tingay},
  C.~M. {Trott}, M.~{Waterson}, R.~B. {Wayth}, R.~L. {Webster}, A.~R.
  {Whitney}, A.~{Williams}, C.~L. {Williams}, and J.~S.~B. {Wyithe}.
\newblock {The Importance of Wide-field Foreground Removal for 21 cm Cosmology:
  A Demonstration with Early MWA Epoch of Reionization Observations}.
\newblock {\em \apj}, 819:8, March 2016.

\bibitem{pober13}
J.~C. {Pober}, A.~R. {Parsons}, J.~E. {Aguirre}, Z.~{Ali}, R.~F. {Bradley},
  C.~L. {Carilli}, D.~{DeBoer}, M.~{Dexter}, N.~E. {Gugliucci}, D.~C. {Jacobs},
  P.~J. {Klima}, D.~{MacMahon}, J.~{Manley}, D.~F. {Moore}, I.~I. {Stefan}, and
  W.~P. {Walbrugh}.
\newblock {Opening the 21 cm Epoch of Reionization Window: Measurements of
  Foreground Isolation with PAPER}.
\newblock {\em \apjl}, 768:L36, May 2013.

\bibitem{procopio17}
P.~{Procopio}, R.~B. {Wayth}, J.~{Line}, C.~M. {Trott}, H.~T. {Intema}, D.~A.
  {Mitchell}, B.~{Pindor}, J.~{Riding}, S.~J. {Tingay}, M.~E. {Bell}, J.~R.
  {Callingham}, K.~S. {Dwarakanath}, B.-Q. {For}, B.~M. {Gaensler}, P.~J.
  {Hancock}, L.~{Hindson}, N.~{Hurley-Walker}, M.~{Johnston-Hollitt}, A.~D.
  {Kapi{\'n}ska}, E.~{Lenc}, B.~{McKinley}, J.~{Morgan}, A.~{Offringa},
  L.~{Staveley-Smith}, C.~{Wu}, and Q.~{Zheng}.
\newblock {A High-Resolution Foreground Model for the MWA EoR1 Field: Model and
  Implications for EoR Power Spectrum Analysis}.
\newblock {\em \pasa}, 34:e033, August 2017.

\bibitem{pupillo15}
G.~{Pupillo}, G.~{Naldi}, G.~{Bianchi}, A.~{Mattana}, J.~{Monari}, F.~{Perini},
  M.~{Poloni}, M.~{Schiaffino}, P.~{Bolli}, A.~{Lingua}, I.~{Aicardi},
  H.~{Bendea}, P.~{Maschio}, M.~{Piras}, G.~{Virone}, F.~{Paonessa},
  Z.~{Farooqui}, A.~{Tibaldi}, G.~{Addamo}, O.~A. {Peverini}, R.~{Tascone}, and
  S.~J. {Wijnholds}.
\newblock {Medicina array demonstrator: calibration and radiation pattern
  characterization using a UAV-mounted radio-frequency source}.
\newblock {\em Experimental Astronomy}, 39:405--421, June 2015.

\bibitem{readhead04}
A.~C.~S. {Readhead}, B.~S. {Mason}, C.~R. {Contaldi}, T.~J. {Pearson}, J.~R.
  {Bond}, S.~T. {Myers}, S.~{Padin}, J.~L. {Sievers}, J.~K. {Cartwright}, M.~C.
  {Shepherd}, D.~{Pogosyan}, S.~{Prunet}, P.~{Altamirano}, R.~{Bustos},
  L.~{Bronfman}, S.~{Casassus}, W.~L. {Holzapfel}, J.~{May}, U.-L. {Pen},
  S.~{Torres}, and P.~S. {Udomprasert}.
\newblock {Extended Mosaic Observations with the Cosmic Background Imager}.
\newblock {\em \apj}, 609:498--512, July 2004.

\bibitem{ryle60}
M.~{Ryle} and A.~{Hewish}.
\newblock {The synthesis of large radio telescopes}.
\newblock {\em \mnras}, 120:220, 1960.

\bibitem{sault96}
R.~J. {Sault}, J.~P. {Hamaker}, and J.~D. {Bregman}.
\newblock {Understanding radio polarimetry. II. Instrumental calibration of an
  interferometer array.}
\newblock {\em \aaps}, 117:149--159, May 1996.

\bibitem{shimwell19}
T.~W. {Shimwell}, C.~{Tasse}, M.~J. {Hardcastle}, A.~P. {Mechev}, W.~L.
  {Williams}, P.~N. {Best}, H.~J.~A. {R{\"o}ttgering}, J.~R. {Callingham},
  T.~J. {Dijkema}, F.~{de Gasperin}, D.~N. {Hoang}, B.~{Hugo}, M.~{Mirmont},
  J.~B.~R. {Oonk}, I.~{Prandoni}, D.~{Rafferty}, J.~{Sabater}, O.~{Smirnov},
  R.~J. {van Weeren}, G.~J. {White}, M.~{Atemkeng}, L.~{Bester},
  E.~{Bonnassieux}, M.~{Br{\"u}ggen}, G.~{Brunetti}, K.~T. {Chy{\.z}y},
  R.~{Cochrane}, J.~E. {Conway}, J.~H. {Croston}, A.~{Danezi}, K.~{Duncan},
  M.~{Haverkorn}, G.~H. {Heald}, M.~{Iacobelli}, H.~T. {Intema}, N.~{Jackson},
  M.~{Jamrozy}, M.~J. {Jarvis}, R.~{Lakhoo}, M.~{Mevius}, G.~K. {Miley},
  L.~{Morabito}, R.~{Morganti}, D.~{Nisbet}, E.~{Orr{\'u}}, S.~{Perkins}, R.~F.
  {Pizzo}, C.~{Schrijvers}, D.~J.~B. {Smith}, R.~{Vermeulen}, M.~W. {Wise},
  L.~{Alegre}, D.~J. {Bacon}, I.~M. {van Bemmel}, R.~J. {Beswick},
  A.~{Bonafede}, A.~{Botteon}, S.~{Bourke}, M.~{Brienza}, G.~{Calistro Rivera},
  R.~{Cassano}, A.~O. {Clarke}, C.~J. {Conselice}, R.~J. {Dettmar},
  A.~{Drabent}, C.~{Dumba}, K.~L. {Emig}, T.~A. {En{\ss}lin}, C.~{Ferrari},
  M.~A. {Garrett}, R.~T. {G{\'e}nova-Santos}, A.~{Goyal}, G.~{G{\"u}rkan},
  C.~{Hale}, J.~J. {Harwood}, V.~{Heesen}, M.~{Hoeft}, C.~{Horellou},
  C.~{Jackson}, G.~{Kokotanekov}, R.~{Kondapally}, M.~{Kunert-Bajraszewska},
  V.~{Mahatma}, E.~K. {Mahony}, S.~{Mandal}, J.~P. {McKean}, A.~{Merloni},
  B.~{Mingo}, A.~{Miskolczi}, S.~{Mooney}, B.~{Nikiel-Wroczy{\'n}ski}, S.~P.
  {O'Sullivan}, J.~{Quinn}, W.~{Reich}, C.~{Roskowi{\'n}ski}, A.~{Rowlinson},
  F.~{Savini}, A.~{Saxena}, D.~J. {Schwarz}, A.~{Shulevski}, S.~S. {Sridhar},
  H.~R. {Stacey}, S.~{Urquhart}, M.~H.~D. {van der Wiel}, E.~{Varenius},
  B.~{Webster}, and A.~{Wilber}.
\newblock {The LOFAR Two-metre Sky Survey. II. First data release}.
\newblock {\em \aap}, 622:A1, February 2019.

\bibitem{smirnov11}
O.~M. {Smirnov}.
\newblock {Revisiting the radio interferometer measurement equation. I. A
  full-sky Jones formalism}.
\newblock {\em \aap}, 527:A106, March 2011.

\bibitem{smirnov11b}
O.~M. {Smirnov}.
\newblock {Revisiting the radio interferometer measurement equation. II.
  Calibration and direction-dependent effects}.
\newblock {\em \aap}, 527:A107, March 2011.

\bibitem{smirnov11c}
O.~M. {Smirnov}.
\newblock {Revisiting the radio interferometer measurement equation. III.
  Addressing direction-dependent effects in 21 cm WSRT observations of 3C 147}.
\newblock {\em \aap}, 527:A108, March 2011.

\bibitem{smirnov15}
O.~M. {Smirnov} and C.~{Tasse}.
\newblock {Radio interferometric gain calibration as a complex optimization
  problem}.
\newblock {\em \mnras}, 449:2668--2684, May 2015.

\bibitem{sokolowski17}
M.~{Sokolowski}, T.~{Colegate}, A.~T. {Sutinjo}, D.~{Ung}, R.~{Wayth},
  N.~{Hurley-Walker}, E.~{Lenc}, B.~{Pindor}, J.~{Morgan}, D.~L. {Kaplan},
  M.~E. {Bell}, J.~R. {Callingham}, K.~S. {Dwarakanath}, B.-Q. {For}, B.~M.
  {Gaensler}, P.~J. {Hancock}, L.~{Hindson}, M.~{Johnston-Hollitt}, A.~D.
  {Kapi{\'n}ska}, B.~{McKinley}, A.~R. {Offringa}, P.~{Procopio},
  L.~{Staveley-Smith}, C.~{Wu}, and Q.~{Zheng}.
\newblock {Calibration and Stokes Imaging with Full Embedded Element Primary
  Beam Model for the Murchison Widefield Array}.
\newblock {\em \pasa}, 34:e062, November 2017.

\bibitem{sullivan12}
I.~S. {Sullivan}, M.~F. {Morales}, B.~J. {Hazelton}, W.~{Arcus}, D.~{Barnes},
  G.~{Bernardi}, F.~H. {Briggs}, J.~D. {Bowman}, J.~D. {Bunton}, R.~J.
  {Cappallo}, B.~E. {Corey}, A.~{Deshpande}, L.~{deSouza}, D.~{Emrich}, B.~M.
  {Gaensler}, R.~{Goeke}, L.~J. {Greenhill}, D.~{Herne}, J.~N. {Hewitt},
  M.~{Johnston-Hollitt}, D.~L. {Kaplan}, J.~C. {Kasper}, B.~B. {Kincaid},
  R.~{Koenig}, E.~{Kratzenberg}, C.~J. {Lonsdale}, M.~J. {Lynch}, S.~R.
  {McWhirter}, D.~A. {Mitchell}, E.~{Morgan}, D.~{Oberoi}, S.~M. {Ord},
  J.~{Pathikulangara}, T.~{Prabu}, R.~A. {Remillard}, A.~E.~E. {Rogers},
  A.~{Roshi}, J.~E. {Salah}, R.~J. {Sault}, N.~{Udaya Shankar}, K.~S.
  {Srivani}, J.~{Stevens}, R.~{Subrahmanyan}, S.~J. {Tingay}, R.~B. {Wayth},
  M.~{Waterson}, R.~L. {Webster}, A.~R. {Whitney}, A.~{Williams}, C.~L.
  {Williams}, and J.~S.~B. {Wyithe}.
\newblock {Fast Holographic Deconvolution: A New Technique for Precision Radio
  Interferometry}.
\newblock {\em \apj}, 759:17, November 2012.

\bibitem{tasse14}
C.~{Tasse}.
\newblock {Nonlinear Kalman filters for calibration in radio interferometry}.
\newblock {\em \aap}, 566:A127, June 2014.

\bibitem{tasse13}
C.~{Tasse}, S.~{van der Tol}, J.~{van Zwieten}, G.~{van Diepen}, and
  S.~{Bhatnagar}.
\newblock {Applying full polarization A-Projection to very wide field of view
  instruments: An imager for LOFAR}.
\newblock {\em \aap}, 553:A105, May 2013.

\bibitem{TMS}
A.~Richard {Thompson}, James~M. {Moran}, and Jr. {Swenson}, George~W.
\newblock {\em {Interferometry and Synthesis in Radio Astronomy, 3rd Edition}}.
\newblock 2017.

\bibitem{thyagarajan13}
N.~{Thyagarajan}, N.~{Udaya Shankar}, R.~{Subrahmanyan}, W.~{Arcus},
  G.~{Bernardi}, J.~D. {Bowman}, F.~{Briggs}, J.~D. {Bunton}, R.~J. {Cappallo},
  B.~E. {Corey}, L.~{deSouza}, D.~{Emrich}, B.~M. {Gaensler}, R.~F. {Goeke},
  L.~J. {Greenhill}, B.~J. {Hazelton}, D.~{Herne}, J.~N. {Hewitt},
  M.~{Johnston-Hollitt}, D.~L. {Kaplan}, J.~C. {Kasper}, B.~B. {Kincaid},
  R.~{Koenig}, E.~{Kratzenberg}, C.~J. {Lonsdale}, M.~J. {Lynch}, S.~R.
  {McWhirter}, D.~A. {Mitchell}, M.~F. {Morales}, E.~H. {Morgan}, D.~{Oberoi},
  S.~M. {Ord}, J.~{Pathikulangara}, R.~A. {Remillard}, A.~E.~E. {Rogers},
  D.~{Anish Roshi}, J.~E. {Salah}, R.~J. {Sault}, K.~S. {Srivani}, J.~B.
  {Stevens}, P.~{Thiagaraj}, S.~J. {Tingay}, R.~B. {Wayth}, M.~{Waterson},
  R.~L. {Webster}, A.~R. {Whitney}, A.~J. {Williams}, C.~L. {Williams}, and
  J.~S.~B. {Wyithe}.
\newblock {A Study of Fundamental Limitations to Statistical Detection of
  Redshifted H I from the Epoch of Reionization}.
\newblock {\em \apj}, 776:6, October 2013.

\bibitem{tingay13}
S.~J. {Tingay}, R.~{Goeke}, J.~D. {Bowman}, D.~{Emrich}, S.~M. {Ord}, D.~A.
  {Mitchell}, M.~F. {Morales}, T.~{Booler}, B.~{Crosse}, R.~B. {Wayth}, C.~J.
  {Lonsdale}, S.~{Tremblay}, D.~{Pallot}, T.~{Colegate}, A.~{Wicenec},
  N.~{Kudryavtseva}, W.~{Arcus}, D.~{Barnes}, G.~{Bernardi}, F.~{Briggs},
  S.~{Burns}, J.~D. {Bunton}, R.~J. {Cappallo}, B.~E. {Corey}, A.~{Deshpande},
  L.~{Desouza}, B.~M. {Gaensler}, L.~J. {Greenhill}, P.~J. {Hall}, B.~J.
  {Hazelton}, D.~{Herne}, J.~N. {Hewitt}, M.~{Johnston-Hollitt}, D.~L.
  {Kaplan}, J.~C. {Kasper}, B.~B. {Kincaid}, R.~{Koenig}, E.~{Kratzenberg},
  M.~J. {Lynch}, B.~{Mckinley}, S.~R. {Mcwhirter}, E.~{Morgan}, D.~{Oberoi},
  J.~{Pathikulangara}, T.~{Prabu}, R.~A. {Remillard}, A.~E.~E. {Rogers},
  A.~{Roshi}, J.~E. {Salah}, R.~J. {Sault}, N.~{Udaya-Shankar},
  F.~{Schlagenhaufer}, K.~S. {Srivani}, J.~{Stevens}, R.~{Subrahmanyan},
  M.~{Waterson}, R.~L. {Webster}, A.~R. {Whitney}, A.~{Williams}, C.~L.
  {Williams}, and J.~S.~B. {Wyithe}.
\newblock {The Murchison Widefield Array: The Square Kilometre Array Precursor
  at Low Radio Frequencies}.
\newblock {\em \pasa}, 30:e007, January 2013.

\bibitem{trott14}
C.~M. {Trott}.
\newblock {Comparison of Observing Modes for Statistical Estimation of the 21
  cm Signal from the Epoch of Reionisation}.
\newblock {\em \pasa}, 31:e026, July 2014.

\bibitem{trott17a}
C.~M. {Trott}, E.~{de Lera Acedo}, R.~B. {Wayth}, N.~{Fagnoni}, A.~T.
  {Sutinjo}, B.~{Wakley}, and C.~I.~B. {Punzalan}.
\newblock {Spectral performance of Square Kilometre Array Antennas - II.
  Calibration performance}.
\newblock {\em \mnras}, 470:455--465, September 2017.

\bibitem{trott17b}
C.~M. {Trott}, C.~H. {Jordan}, S.~G. {Murray}, B.~{Pindor}, D.~A. {Mitchell},
  R.~B. {Wayth}, J.~{Line}, B.~{McKinley}, A.~{Beardsley}, J.~{Bowman},
  F.~{Briggs}, B.~J. {Hazelton}, J.~{Hewitt}, D.~{Jacobs}, M.~F. {Morales},
  J.~C. {Pober}, S.~{Sethi}, U.~{Shankar}, R.~{Subrahmanyan}, M.~{Tegmark},
  S.~J. {Tingay}, R.~L. {Webster}, and J.~S.~B. {Wyithe}.
\newblock {Assessment of Ionospheric Activity Tolerances for Epoch of
  Reionization Science with the Murchison Widefield Array}.
\newblock {\em \apj}, 867:15, November 2018.

\bibitem{trott16}
C.~M. {Trott} and R.~B. {Wayth}.
\newblock {Spectral Calibration Requirements of Radio Interferometers for Epoch
  of Reionisation Science with the SKA}.
\newblock {\em \pasa}, 33:e019, May 2016.

\bibitem{vaneck18}
C.~L. {Van Eck}, M.~{Haverkorn}, M.~I.~R. {Alves}, R.~{Beck}, P.~{Best},
  E.~{Carretti}, K.~T. {Chy{\.z}y}, J.~S. {Farnes}, K.~{Ferri{\`e}re}, M.~J.
  {Hardcastle}, G.~{Heald}, C.~{Horellou}, M.~{Iacobelli}, V.~{Jeli{\'c}},
  D.~D. {Mulcahy}, S.~P. {O'Sullivan}, I.~M. {Polderman}, W.~{Reich}, C.~J.
  {Riseley}, H.~{R{\"o}ttgering}, D.~H.~F.~M. {Schnitzeler}, T.~W. {Shimwell},
  V.~{Vacca}, J.~{Vink}, and G.~J. {White}.
\newblock {Polarized point sources in the LOFAR Two-meter Sky Survey: A
  preliminary catalog}.
\newblock {\em \aap}, 613:A58, June 2018.

\bibitem{vanhaarlem13}
M.~P. {van Haarlem}, M.~W. {Wise}, A.~W. {Gunst}, G.~{Heald}, J.~P. {McKean},
  J.~W.~T. {Hessels}, A.~G. {de Bruyn}, R.~{Nijboer}, J.~{Swinbank},
  R.~{Fallows}, M.~{Brentjens}, A.~{Nelles}, R.~{Beck}, H.~{Falcke},
  R.~{Fender}, J.~{H{\"o}randel}, L.~V.~E. {Koopmans}, G.~{Mann}, G.~{Miley},
  H.~{R{\"o}ttgering}, B.~W. {Stappers}, R.~A.~M.~J. {Wijers}, S.~{Zaroubi},
  M.~{van den Akker}, A.~{Alexov}, J.~{Anderson}, K.~{Anderson}, A.~{van
  Ardenne}, M.~{Arts}, A.~{Asgekar}, I.~M. {Avruch}, F.~{Batejat},
  L.~{B{\"a}hren}, M.~E. {Bell}, M.~R. {Bell}, I.~{van Bemmel}, P.~{Bennema},
  M.~J. {Bentum}, G.~{Bernardi}, P.~{Best}, L.~{B{\^i}rzan}, A.~{Bonafede},
  A.-J. {Boonstra}, R.~{Braun}, J.~{Bregman}, F.~{Breitling}, R.~H. {van de
  Brink}, J.~{Broderick}, P.~C. {Broekema}, W.~N. {Brouw}, M.~{Br{\"u}ggen},
  H.~R. {Butcher}, W.~{van Cappellen}, B.~{Ciardi}, T.~{Coenen}, J.~{Conway},
  A.~{Coolen}, A.~{Corstanje}, S.~{Damstra}, O.~{Davies}, A.~T. {Deller}, R.-J.
  {Dettmar}, G.~{van Diepen}, K.~{Dijkstra}, P.~{Donker}, A.~{Doorduin},
  J.~{Dromer}, M.~{Drost}, A.~{van Duin}, J.~{Eisl{\"o}ffel}, J.~{van Enst},
  C.~{Ferrari}, W.~{Frieswijk}, H.~{Gankema}, M.~A. {Garrett}, F.~{de
  Gasperin}, M.~{Gerbers}, E.~{de Geus}, J.-M. {Grie{\ss}meier}, T.~{Grit},
  P.~{Gruppen}, J.~P. {Hamaker}, T.~{Hassall}, M.~{Hoeft}, H.~A. {Holties},
  A.~{Horneffer}, A.~{van der Horst}, A.~{van Houwelingen}, A.~{Huijgen},
  M.~{Iacobelli}, H.~{Intema}, N.~{Jackson}, V.~{Jelic}, A.~{de Jong},
  E.~{Juette}, D.~{Kant}, A.~{Karastergiou}, A.~{Koers}, H.~{Kollen}, V.~I.
  {Kondratiev}, E.~{Kooistra}, Y.~{Koopman}, A.~{Koster}, M.~{Kuniyoshi},
  M.~{Kramer}, G.~{Kuper}, P.~{Lambropoulos}, C.~{Law}, J.~{van Leeuwen},
  J.~{Lemaitre}, M.~{Loose}, P.~{Maat}, G.~{Macario}, S.~{Markoff},
  J.~{Masters}, R.~A. {McFadden}, D.~{McKay-Bukowski}, H.~{Meijering},
  H.~{Meulman}, M.~{Mevius}, E.~{Middelberg}, R.~{Millenaar}, J.~C.~A.
  {Miller-Jones}, R.~N. {Mohan}, J.~D. {Mol}, J.~{Morawietz}, R.~{Morganti},
  D.~D. {Mulcahy}, E.~{Mulder}, H.~{Munk}, L.~{Nieuwenhuis}, R.~{van
  Nieuwpoort}, J.~E. {Noordam}, M.~{Norden}, A.~{Noutsos}, A.~R. {Offringa},
  H.~{Olofsson}, A.~{Omar}, E.~{Orr{\'u}}, R.~{Overeem}, H.~{Paas},
  M.~{Pandey-Pommier}, V.~N. {Pandey}, R.~{Pizzo}, A.~{Polatidis},
  D.~{Rafferty}, S.~{Rawlings}, W.~{Reich}, J.-P. {de Reijer}, J.~{Reitsma},
  G.~A. {Renting}, P.~{Riemers}, E.~{Rol}, J.~W. {Romein}, J.~{Roosjen},
  M.~{Ruiter}, A.~{Scaife}, K.~{van der Schaaf}, B.~{Scheers}, P.~{Schellart},
  A.~{Schoenmakers}, G.~{Schoonderbeek}, M.~{Serylak}, A.~{Shulevski},
  J.~{Sluman}, O.~{Smirnov}, C.~{Sobey}, H.~{Spreeuw}, M.~{Steinmetz}, C.~G.~M.
  {Sterks}, H.-J. {Stiepel}, K.~{Stuurwold}, M.~{Tagger}, Y.~{Tang},
  C.~{Tasse}, I.~{Thomas}, S.~{Thoudam}, M.~C. {Toribio}, B.~{van der Tol},
  O.~{Usov}, M.~{van Veelen}, A.-J. {van der Veen}, S.~{ter Veen}, J.~P.~W.
  {Verbiest}, R.~{Vermeulen}, N.~{Vermaas}, C.~{Vocks}, C.~{Vogt}, M.~{de Vos},
  E.~{van der Wal}, R.~{van Weeren}, H.~{Weggemans}, P.~{Weltevrede},
  S.~{White}, S.~J. {Wijnholds}, T.~{Wilhelmsson}, O.~{Wucknitz},
  S.~{Yatawatta}, P.~{Zarka}, A.~{Zensus}, and J.~{van Zwieten}.
\newblock {LOFAR: The LOw-Frequency ARray}.
\newblock {\em \aap}, 556:A2, August 2013.

\bibitem{vanweeren16}
R.~J. {van Weeren}, W.~L. {Williams}, M.~J. {Hardcastle}, T.~W. {Shimwell},
  D.~A. {Rafferty}, J.~{Sabater}, G.~{Heald}, S.~S. {Sridhar}, T.~J. {Dijkema},
  G.~{Brunetti}, M.~{Br{\"u}ggen}, F.~{Andrade-Santos}, G.~A. {Ogrean},
  H.~J.~A. {R{\"o}ttgering}, W.~A. {Dawson}, W.~R. {Forman}, F.~{de Gasperin},
  C.~{Jones}, G.~K. {Miley}, L.~{Rudnick}, C.~L. {Sarazin}, A.~{Bonafede},
  P.~N. {Best}, L.~{B{\^i}rzan}, R.~{Cassano}, K.~T. {Chy{\.z}y}, J.~H.
  {Croston}, T.~{Ensslin}, C.~{Ferrari}, M.~{Hoeft}, C.~{Horellou}, M.~J.
  {Jarvis}, R.~P. {Kraft}, M.~{Mevius}, H.~T. {Intema}, S.~S. {Murray},
  E.~{Orr{\'u}}, R.~{Pizzo}, A.~{Simionescu}, A.~{Stroe}, S.~{van der Tol}, and
  G.~J. {White}.
\newblock {LOFAR Facet Calibration}.
\newblock {\em \apjs}, 223:2, March 2016.

\bibitem{vedantham12}
H.~{Vedantham}, N.~{Udaya Shankar}, and R.~{Subrahmanyan}.
\newblock {Imaging the Epoch of Reionization: Limitations from Foreground
  Confusion and Imaging Algorithms}.
\newblock {\em \apj}, 745:176, February 2012.

\bibitem{vedantham16}
H.~K. {Vedantham} and L.~V.~E. {Koopmans}.
\newblock {Scintillation noise power spectrum and its impact on high-redshift
  21-cm observations}.
\newblock {\em \mnras}, 458:3099--3117, May 2016.

\bibitem{wayth18}
R.~B. {Wayth}, S.~J. {Tingay}, C.~M. {Trott}, D.~{Emrich},
  M.~{Johnston-Hollitt}, B.~{McKinley}, B.~M. {Gaensler}, A.~P. {Beardsley},
  T.~{Booler}, B.~{Crosse}, T.~M.~O. {Franzen}, L.~{Horsley}, D.~L. {Kaplan},
  D.~{Kenney}, M.~F. {Morales}, D.~{Pallot}, G.~{Sleap}, K.~{Steele},
  M.~{Walker}, A.~{Williams}, C.~{Wu}, I.~H. {Cairns}, M.~D. {Filipovic},
  S.~{Johnston}, T.~{Murphy}, P.~{Quinn}, L.~{Staveley-Smith}, R.~{Webster},
  and J.~S.~B. {Wyithe}.
\newblock {The Phase II Murchison Widefield Array: Design overview}.
\newblock {\em \pasa}, 35, November 2018.

\bibitem{white99}
M.~{White}, J.~E. {Carlstrom}, M.~{Dragovan}, and W.~L. {Holzapfel}.
\newblock {Interferometric Observation of Cosmic Microwave Background
  Anisotropies}.
\newblock {\em \apj}, 514:12--24, March 1999.

\bibitem{wieringa92}
M.~H. {Wieringa}.
\newblock {An investigation of the telescope based calibration methods
  'redundancy' and 'self-cal'}.
\newblock {\em Experimental Astronomy}, 2:203--225, 1992.

\bibitem{wijnholds16}
S.~J. {Wijnholds}, T.~L. {Grobler}, and O.~M. {Smirnov}.
\newblock {Calibration artefacts in radio interferometry - II. Ghost patterns
  for irregular arrays}.
\newblock {\em \mnras}, 457:2331--2354, April 2016.

\bibitem{yatawatta15}
S.~{Yatawatta}.
\newblock {Distributed radio interferometric calibration}.
\newblock {\em \mnras}, 449:4506--4514, June 2015.

\bibitem{yatawatta13}
S.~{Yatawatta}, A.~G. {de Bruyn}, M.~A. {Brentjens}, P.~{Labropoulos}, V.~N.
  {Pandey}, S.~{Kazemi}, S.~{Zaroubi}, L.~V.~E. {Koopmans}, A.~R. {Offringa},
  V.~{Jeli{\'c}}, O.~{Martinez Rubi}, V.~{Veligatla}, S.~J. {Wijnholds}, W.~N.
  {Brouw}, G.~{Bernardi}, B.~{Ciardi}, S.~{Daiboo}, G.~{Harker}, G.~{Mellema},
  J.~{Schaye}, R.~{Thomas}, H.~{Vedantham}, E.~{Chapman}, F.~B. {Abdalla},
  A.~{Alexov}, J.~{Anderson}, I.~M. {Avruch}, F.~{Batejat}, M.~E. {Bell}, M.~R.
  {Bell}, M.~{Bentum}, P.~{Best}, A.~{Bonafede}, J.~{Bregman}, F.~{Breitling},
  R.~H. {van de Brink}, J.~W. {Broderick}, M.~{Br{\"u}ggen}, J.~{Conway},
  F.~{de Gasperin}, E.~{de Geus}, S.~{Duscha}, H.~{Falcke}, R.~A. {Fallows},
  C.~{Ferrari}, W.~{Frieswijk}, M.~A. {Garrett}, J.~M. {Griessmeier}, A.~W.
  {Gunst}, T.~E. {Hassall}, J.~W.~T. {Hessels}, M.~{Hoeft}, M.~{Iacobelli},
  E.~{Juette}, A.~{Karastergiou}, V.~I. {Kondratiev}, M.~{Kramer},
  M.~{Kuniyoshi}, G.~{Kuper}, J.~{van Leeuwen}, P.~{Maat}, G.~{Mann}, J.~P.
  {McKean}, M.~{Mevius}, J.~D. {Mol}, H.~{Munk}, R.~{Nijboer}, J.~E. {Noordam},
  M.~J. {Norden}, E.~{Orru}, H.~{Paas}, M.~{Pandey-Pommier}, R.~{Pizzo}, A.~G.
  {Polatidis}, W.~{Reich}, H.~J.~A. {R{\"o}ttgering}, J.~{Sluman},
  O.~{Smirnov}, B.~{Stappers}, M.~{Steinmetz}, M.~{Tagger}, Y.~{Tang},
  C.~{Tasse}, S.~{ter Veen}, R.~{Vermeulen}, R.~J. {van Weeren}, M.~{Wise},
  O.~{Wucknitz}, and P.~{Zarka}.
\newblock {Initial deep LOFAR observations of epoch of reionization windows. I.
  The north celestial pole}.
\newblock {\em \aap}, 550:A136, February 2013.

\bibitem{zaroubi12}
S.~{Zaroubi}, A.~G. {de Bruyn}, G.~{Harker}, R.~M. {Thomas}, P.~{Labropolous},
  V.~{Jeli{\'c}}, L.~V.~E. {Koopmans}, M.~A. {Brentjens}, G.~{Bernardi},
  B.~{Ciardi}, S.~{Daiboo}, S.~{Kazemi}, O.~{Martinez-Rubi}, G.~{Mellema},
  A.~R. {Offringa}, V.~N. {Pandey}, J.~{Schaye}, V.~{Veligatla},
  H.~{Vedantham}, and S.~{Yatawatta}.
\newblock {Imaging neutral hydrogen on large scales during the Epoch of
  Reionization with LOFAR}.
\newblock {\em \mnras}, 425:2964--2973, October 2012.

\bibitem{zheng14}
H.~{Zheng}, M.~{Tegmark}, V.~{Buza}, J.~S. {Dillon}, H.~{Gharibyan},
  J.~{Hickish}, E.~{Kunz}, A.~{Liu}, J.~{Losh}, A.~{Lutomirski}, S.~{Morrison},
  S.~{Narayanan}, A.~{Perko}, D.~{Rosner}, N.~{Sanchez}, K.~{Schutz}, S.~M.
  {Tribiano}, M.~{Valdez}, H.~{Yang}, K.~Z. {Adami}, I.~{Zelko}, K.~{Zheng},
  R.~P. {Armstrong}, R.~F. {Bradley}, M.~R. {Dexter}, A.~{Ewall-Wice},
  A.~{Magro}, M.~{Matejek}, E.~{Morgan}, A.~R. {Neben}, Q.~{Pan}, R.~F.
  {Penna}, C.~M. {Peterson}, M.~{Su}, J.~{Villasenor}, C.~L. {Williams}, and
  Y.~{Zhu}.
\newblock {MITEoR: a scalable interferometer for precision 21 cm cosmology}.
\newblock {\em \mnras}, 445:1084--1103, December 2014.

\bibitem{zheng17}
H.~{Zheng}, M.~{Tegmark}, J.~S. {Dillon}, A.~{Liu}, A.~R. {Neben}, S.~M.
  {Tribiano}, R.~F. {Bradley}, V.~{Buza}, A.~{Ewall-Wice}, H.~{Gharibyan},
  J.~{Hickish}, E.~{Kunz}, J.~{Losh}, A.~{Lutomirski}, E.~{Morgan},
  S.~{Narayanan}, A.~{Perko}, D.~{Rosner}, N.~{Sanchez}, K.~{Schutz},
  M.~{Valdez}, J.~{Villasenor}, H.~{Yang}, K.~{Zarb Adami}, I.~{Zelko}, and
  K.~{Zheng}.
\newblock {Brute-force mapmaking with compact interferometers: a MITEoR
  northern sky map from 128 to 175 MHz}.
\newblock {\em \mnras}, 465:2901--2915, March 2017.

\end{thebibliography}
